\documentclass[
	paper=a4,%
	11pt,american,pagesize,%
	DIV=12,headinclude=true,headlines=0,footinclude=true,footlines=-5,twoside=semi%
]{scrartcl}
\def\version{arxiv}

\def\draftmode{false}

\usepackage{hyperref}

\RequirePackage{etex}
\RequirePackage{fixltx2e}

\makeatletter

\usepackage{lmodern}
\usepackage[utf8]{inputenc}
\usepackage[T1]{fontenc}
\usepackage{babel}
\usepackage{amsmath,amssymb,amsfonts,mathtools,colonequals}
\mathtoolsset{mathic=true}
\usepackage{calc,relsize,xifthen,xcolor,chngcntr,xspace,adjustbox,pbox}
\usepackage{array,multirow,multicol,booktabs,rotating}
\usepackage{enumitem}
\usepackage{placeins,needspace}
\usepackage{mleftright}\mleftright %
\usepackage{url}

\usepackage{xfrac,dsfont,bm} %
\usepackage{stmaryrd} %

\newcommand{\ifarxiv}[2]{\ifthenelse{\equal{\version}{arxiv}}{#1}{#2}}
\newcommand{\ifdraft}[2]{\ifthenelse{\equal{\draftmode}{true}}{#1}{#2}}
\newcommand{\ifsiam}[2]{\ifthenelse{\equal{\version}{proceedings-siam}}{#1}{#2}}
\newcommand{\ifsubmission}[2]{\ifthenelse{\equal{\version}{submission}}{#1}{#2}}

	\usepackage{microtype}

\usepackage{wref}
\usepackage{needhspace}

\ifsiam{}{
	\setlength\parindent{1.5em}
	\usepackage[headsepline]{scrlayer-scrpage}
	\pagestyle{scrheadings}
	\clearscrheadfoot
	\AtBeginDocument{%
		\manualmark%
		\markleft{\mytitle}%
		\automark*[section]{}%
	}
	\ohead{\pagemark}
	\ihead{\headmark}
	\addtokomafont{caption}{\sffamily\small}
	\addtokomafont{captionlabel}{\sffamily\textbf}
	\setcapmargin{2em}
}
\ifsubmission{
	\setcapmargin{.75em}
	\setcapindent{0pt}
}{}
\ifarxiv{
	\setcapmargin{.75em}
	\setcapindent{0pt}
}{}

\usepackage{tikz}
\usetikzlibrary{babel,quotes}
\usetikzlibrary{patterns}
\usetikzlibrary{positioning,external,calc,graphs}
\usetikzlibrary{decorations.pathreplacing,shapes.multipart,shapes.geometric}
\usetikzlibrary{fit,arrows.meta}
\usepackage{ifluatex}
\ifluatex
	\usetikzlibrary{graphdrawing}
	\usegdlibrary{trees,layered}
\else \fi

\tikzsetexternalprefix{figs/externalized/}
\tikzset{external/export=false} %

\newcommand{\externalizedpicture}[1]{%
	\tikzset{external/export=true}%
	\tikzsetnextfilename{#1}%
}

\pgfdeclarelayer{foreground}
\pgfdeclarelayer{background}
\pgfsetlayers{background,main,foreground}

\tikzset{%
	every picture/.style={
		>=stealth,
		font={\sffamily\small},
	},
}

\newcommand\currentcoordinate{\the\tikz@lastxsaved,\the\tikz@lastysaved}
\newcommand\currentx{\the\tikz@lastxsaved}
\newcommand\currenty{\the\tikz@lastysaved}

\usepackage{pgfplots}
\pgfplotsset{compat=1.12}
\ifsiam{
	\usepackage[margin=.75em,font={small,rm},labelfont={bf}]{caption}
	\DeclareCaptionStyle{ruled}{%
		format=plain,%
		labelfont=bf,%
		labelsep=period,%
		strut=on,%
		margin=.3em,%
		font={small,rm}%
	}
}{}

\usepackage{clrscode3e}

\usepackage{float}
\floatstyle{ruled}
\newfloat{algorithm}{tbp}{loa}
\floatname{algorithm}{Algorithm}
\hypersetup{
	final,
	unicode=true, 
	bookmarks=true,
	bookmarksnumbered=true,
	bookmarksdepth=2,
	bookmarksopen=true,
	breaklinks=true,
	hidelinks,
}

\usepackage{attachfile2}

\usepackage[amsmath,hyperref,thmmarks]{ntheorem}

\theorembodyfont{\slshape}
\theoremseparator{:}
\newtheoremstyle{proofstyle}%
  {\item[\theorem@headerfont\hskip\labelsep ##1\theorem@separator]}%
  {\item[\theorem@headerfont\hskip\labelsep ##1 of ##3\theorem@separator]}

\theoremstyle{break}
\ifsubmission{
	\theorempreskip{.59\baselineskip}
	\theorempostskip{.5\baselineskip}
}{
	\theorempreskip{2\topsep}
}
\ifsiam{
	\renewtheorem{theorem}{Theorem}[section]
}{
	\newtheorem{theorem}{Theorem}[section]
}

\theoremstyle{plain}
\theorempreskip{\topsep}

\ifsiam{
	\renewtheorem{lemma}[theorem]{Lemma}
	\renewtheorem{corollary}[theorem]{Corollary}
	\renewtheorem{proposition}[theorem]{Proposition}
}{
	\newtheorem{lemma}[theorem]{Lemma}

}

\theoremstyle{plain}
\theorembodyfont{\upshape}

\newtheorem{remark}[theorem]{Remark}

\theoremsymbol{\raisebox{-.25ex}{$\Box$}}
\qedsymbol{\raisebox{-.25ex}{$\Box$}}

\theoremstyle{proofstyle}
\ifsiam{
	\renewtheorem{proof}{Proof}
}{
	\newtheorem{proof}{Proof}
}
\vfuzz \hfuzz	

\ifarxiv{
	\raggedbottom
}{}

\allowdisplaybreaks[3]

\pgfplotsset{
	legend cell align=left,
	every axis legend/.append style={%
		fill=backgroundshading!50!white,
		rounded corners=1pt,
	},
	xlabel near ticks,
	ylabel near ticks,
}
\pgfplotsset{
	every y tick scale label/.style={
		at={(-.03,1)},yshift=-1ex,anchor=south east,inner sep=0pt
	}
}

\def\mathematicaPlotAddPlot{%
	\@ifstar\@mathematicaPlotAddPlot\@@mathematicaPlotAddPlot%
}
\newcommand\@@mathematicaPlotAddPlot[4][]{%
	\addplot+[#1] table[x=#2,y=#3] {\tablefile} ;
	\ifthenelse{\equal{#4}{}}{}{\addlegendentry{#4}}
}
\newcommand\@mathematicaPlotAddPlot[4][]{%
	\addplot[#1] table[x=#2,y=#3] {\tablefile} ;
	\ifthenelse{\equal{#4}{}}{}{\addlegendentry{#4}}
}

\pgfplotscreateplotcyclelist{coloredlineplots}{%
	thick,headingscolor\\%
	thick,black\\%
	ultra thick,headingscolor,densely dotted\\%
	very thick,black,densely dashed\\%
	semithick,headingscolor,dashdotted\\%
	thin,black,loosely dashed\\%
}

\pgfplotscreateplotcyclelist{coloreddiscreteplots}{%
	mark=*,headingscolor,every mark/.append style={fill=headingscolor!50}%
\\%
	mark=square*,every mark/.append style={fill=white}%
\\%
	densely dashed,mark=square*,headingscolor,%
		every mark/.append style={solid,fill=red!10}%
\\%
	densely dashed,mark=*,%
		every mark/.append style={solid,fill=black!50}%
\\%
	mark=diamond*,every mark/.append style={%
		semithick,solid,fill=headingscolor!30!black},%
	headingscolor,very thick,densely dotted,%
\\%
}

\newdimen\makeboxdimen
\newcommand\makeboxlike[3][l]{%
\setbox0=\hbox{#2}%
\global\makeboxdimen=\wd0%
\setbox1=\hbox{\makebox[\makeboxdimen][#1]{%
\makebox[0pt][#1]{#3}%
}}%
\ht1=\ht0%
\dp1=\dp0%
\box1%
}

\newcommand\plaincenter[1]{%
	\mbox{}\hfill#1\hfill\mbox{}%
}

\newcounter{inlineenum}
\newenvironment{inlineenumerate}{%
	\setcounter{inlineenum}{0}%
	\def\item{%
		\stepcounter{inlineenum}%
		\arabic{inlineenum})~%
	}%
}{%
}

\newcommand*\ie{\mbox{i.\hspace{.2ex}e.}}
\newcommand*\eg{\mbox{e.\hspace{.2ex}g.}}
\newcommand*\iid{\mbox{i.\hspace{.2ex}i.\hspace{.2ex}d.}\xspace}

\newcommand*\wrt{\mbox{w.\hspace{.2ex}r.\hspace{.2ex}t.}\xspace}

\newcommand*\withoutlossofgenerality{\mbox{w.\hspace{.23ex}l.\hspace{.2ex}o.\hspace{.17ex}g.}\xspace}

\newcommand\literalquote[3]{%
	{%
		\def\ellipsis{[\,\dots]\xspace}%
		\def\comment##1{[##1]}%
		\def\cite##1{##1}%
		\edef\citekey{#2}%
		{\itshape``\ignorespaces#1\unskip''}
		(%
			\ifthenelse{\equal{#3}{}}{%
			\cite{\citekey}%
			}{%
			\cite{\citekey}%
			, p.\,#3%
			}%
		)%
	}%
	\xspace%
}
\newcommand\literalquotegerman[3]{%
	{%
		\def\ellipsis{[\,\dots]\xspace}%
		\def\comment##1{[##1]}%
		\def\cite##1{##1}%
		\edef\citekey{#2}%
		{\selectlanguage{ngerman}%
		\itshape,,\ignorespaces#1\unskip``}
		(%
			\ifthenelse{\equal{#3}{}}{%
			\cite{\citekey}%
			}{%
			\cite{\citekey}%
			, p.\,#3%
			}%
		)%
	}%
	\xspace%
}
\newcommand\literalquotep[3]{%
	{%
		\def\ellipsis{[\,\dots]\xspace}%
		\def\comment##1{[##1]}%
		\def\cite##1{##1}%
		\edef\citekey{#2}%
		{\itshape``\ignorespaces#1\unskip''}
		(%
			\ifthenelse{\equal{#3}{}}{%
			\cite{\citekey}%
			}{%
			\cite{\citekey}%
			, p.\,#3%
			}%
		)%
	}%
	\xspace%
}

\newcommand{\ESymbol}{\mathbb{E}}

\newcommand\R{\mathbb R}
\newcommand\N{\mathbb N}
\newcommand\Z{\mathbb Z}
\newcommand\Q{\mathbb Q}
\renewcommand\C{\mathbb{C}}
\newcommand\Oh{O}
\def\.{\mskip1mu}

\DeclareMathOperator\ld{ld}

\newcommand{\ProbSymbol}{\ensuremath{\mathbb{P}}}
\providecommand{\given}{}
\DeclarePairedDelimiterXPP\Prob[1]{\ProbSymbol}[]{}{%
	\renewcommand\given{\nonscript\:\delimsize\vert\nonscript\:\mathopen{}}%
	#1%
}
\DeclarePairedDelimiterXPP\E[1]{\ESymbol}[]{}{%
	\renewcommand\given{\nonscript\:\delimsize\vert\nonscript\:\mathopen{}}%
	#1%
}
\DeclarePairedDelimiterXPP\Eover[2]{\ESymbol_{#1}}[]{}{%
	\renewcommand\given{\nonscript\:\delimsize\vert\nonscript\:\mathopen{}}%
	#2%
}
\providecommand{\Prob}{} %
\providecommand{\E}{} %
\providecommand{\Eover}{} %

\newcommand\total[1]{\Sigma #1}

\newcommand\ui[2]{#1^{\smash{(}#2\smash{)}}}

\usepackage{fixmath}
\newcommand{\vect}[1]{\boldsymbol{\mathbold{#1}}}

\newcommand\eqdist{	
	\mathchoice{%
		\mathrel{\overset{\raisebox{0ex}{$\scriptstyle \cal D$}}=}%
	}{%
		\mathrel{\like{=}{%
			\overset{\raisebox{-1ex}{$\scriptscriptstyle \cal D$}}=%
		}}%
	}{%
		\mathrel{\overset{\cal D}=}%
	}{%
		\mathrel{\overset{\cal D}=}%
	}%
}

\newcommand\ppe{\phantom{=}}

\newcommand\like[3][c]{%
	\mathchoice{%
		\makeboxlike[#1]{%
			\ensuremath{\displaystyle\relax#2}%
		}{%
			\ensuremath{\displaystyle\relax#3}%
		}%
	}{%
		\makeboxlike[#1]{%
			\ensuremath{\textstyle\relax#2}%
		}{%
			\ensuremath{\textstyle\relax#3}%
		}%
	}{%
		\makeboxlike[#1]{%
			\ensuremath{\scriptstyle\relax#2}%
		}{%
			\ensuremath{\scriptstyle\relax#3}%
		}%
	}{%
		\makeboxlike[#1]{%
			\ensuremath{\scriptscriptstyle\relax#2}%
		}{%
			\ensuremath{\scriptscriptstyle\relax#3}%
		}%
	}
}

\newcommand\uniform{\mathcal U}

\newcommand\multinomial{\mathrm{Mult}}
\newcommand\binomial{\mathrm{Bin}}
\newcommand\dirichlet{\mathrm{Dir}}
\newcommand\betadist{\mathrm{Beta}}

\newcommand\betaBinomial{\mathrm{BetaBin}}
\newcommand\dirichletMultinomial{\mathrm{DirMult}}

\DeclarePairedDelimiterXPP\distFromWeights[1]{\mathcal D}(){}{#1}
\providecommand\distFromWeights{} %

\newcommand\BetaFun{\mathrm B}

\renewcommand\given{\mathbin{\mid}}
\newcommand\harm[1]{\ensuremath{H_{#1}}}

\newcommand\ce{\colonequals}

\newcommand{\surroundedmath}[3]{%
	\mathchoice{%
		#1{#2{#3}#2}%
	}{%
		#1{#3}%
	}{%
		#1{#3}%
	}{%
		#1{#3}%
	}%
}

\newcommand\rel[1]{\surroundedmath{\mathrel}{\:}{#1}}
\newcommand\wrel[1]{\surroundedmath{\mathrel}{\;}{#1}}
\newcommand\wwrel[1]{\surroundedmath{\mathrel}{\;\;}{#1}}
\newcommand\bin[1]{\surroundedmath{\mathbin}{\:}{#1}}
\newcommand\wbin[1]{\surroundedmath{\mathbin}{\;}{#1}}
\newcommand\wwbin[1]{\surroundedmath{\mathbin}{\;\;}{#1}}

\newcommand{\eqwithref}[2][c]{%
	\relwithref[#1]{#2}{=}%
}
\newcommand{\relwithref}[3][c]{%
	\mathrel{\underset{\mathclap{\makebox[\widthof{$=$}][#1]{\scriptsize\wref{#2}}}}{#3}}%
}
\newcommand{\relwithtext}[3][c]{%
	\mathrel{\underset{\mathclap{\makebox[\widthof{$=$}][#1]{\scriptsize#2}}}{#3}}%
}
\newcommand\relwithtworefs[3]{%
	\relwithtext[c]{\llap{\wref{#1},~}\wref{#2}}{#3}%
}
\newcommand\relwiththreerefs[4]{%
	\relwithtext[c]{\llap{\wref{#1},~\wref{#2},~}\wref{#3}}{#4}%
}

\newcommand\indicator[1]{\indicatornobraces{\{#1\}}}
\newcommand\indicatornobraces[1]{\mathds 1_{#1}}
\newcommand\characteristicvector[2][]{%
	\ifthenelse{\equal{#1}{}}{%
		\mathds1_{#2}%
	}{%
		\ui{\mathds1_{#2}}{#1}%
	}
}

\newcommand\entropy[1][]{%
	\ensuremath{%
		\mathcal H_{#1}%
	}\xspace%
}

\newcommand\discreteEntropy[1][]{%
	\ensuremath{
		\entropy[] \ifthenelse{\equal{#1}{}}{ }{ (#1) }
	}\xspace%
}

\newcommand\constEntropy[1][]{%
	\ensuremath{
		\entropy[] \ifthenelse{\equal{#1}{}}{ }{ (#1) }
	}\xspace%
}

\newcommand\contentropy[1][]{%
	\ensuremath{%
		\entropy[\ln]%
		\ifthenelse{\equal{#1}{}}{ }{ (#1) }%
	}\xspace%
}

\newcommand\binaryEntropy[1][]{%
	\ensuremath{
		\entropy[\ld] \ifthenelse{\equal{#1}{}}{ }{ (#1) }%
	}\xspace%
}

\newcommand\convAS{	
	\mathchoice{%
		\mathrel{\overset{\raisebox{0ex}{\text{a.s.}}}\longrightarrow}%
	}{%
		\mathrel{\like{\longrightarrow}{%
			\overset{\raisebox{-1ex}{\kern-1pt\scriptsize\text{a.s.}}}\longrightarrow%
		}}%
	}{%
		\mathrel{\overset{\text{a.s.}}\longrightarrow}%
	}{%
		\mathrel{\overset{\text{a.s.}}\longrightarrow}%
	}%
}

\newcommand\convD{	
	\mathchoice{%
		\mathrel{\overset{\raisebox{0ex}{$\scriptstyle \cal D$}}\longrightarrow}%
	}{%
		\mathrel{\like{\longrightarrow}{%
			\overset{\raisebox{-1ex}{$\scriptscriptstyle \cal D$}}\longrightarrow%
		}}%
	}{%
		\mathrel{\overset{\cal D}\longrightarrow}%
	}{%
		\mathrel{\overset{\cal D}\longrightarrow}%
	}%
}

\DeclareRobustCommand\arrayA{%
	\ensuremath{\mathchoice{%
		\smash{\raisebox{-.2pt}{\scalebox{1.25}[1.18]{$\mathtt{A}$}}}%
	}{%
		\smash{\raisebox{-.2pt}{\scalebox{1.25}[1.18]{$\mathtt{A}$}}}%
	}{%
		\smash{\raisebox{-.2pt}{\scalebox{1.25}[1.18]{$\scriptstyle\mathtt{A}$}}}%
	}{%
		\smash{\raisebox{-.2pt}{\scalebox{1.25}[1.18]{$\scriptscriptstyle\mathtt{A}$}}}%
	}}\xspace%
}
\providecommand\arrayA

\usepackage{manfnt}

\DeclarePairedDelimiterXPP\dirichletExpectation[2]{\ESymbol_{D(#2)}}[]{}{%
	\renewcommand\given{\nonscript\:\delimsize\vert\nonscript\:\mathopen{}}%
	#1%
}
\providecommand\dirichletExpectation{} %

\let\oldalign\align
\let\endoldalign\endalign

\newcommand*{\numberthis}{\stepcounter{equation}\tag{\theequation}}

\renewenvironment{align}{%
	\begingroup%
	\let\oldhalign\halign
	\def\halign{%
		\let\oldbreak\\%
		\def\nonnumberbreak{\nonumber\oldbreak*}%
		\def\\{%
			\@ifstar{\nonnumberbreak}{\oldbreak}%
		}%
		\oldhalign%
	}
	\oldalign%
}{%
	\endoldalign%
	\endgroup%
}

\let\oldparagraph\paragraph
\renewcommand\paragraph{%
    \@ifstar{\myparagraphStar}{\myparagraphNoStar}%
}
\newcommand\myparagraphStar[1]{%
	\oldparagraph*{#1.}%
}
\newcommand\myparagraphNoStar[2][]{%
	\ifthenelse{\equal{#1}{}}{%
		\oldparagraph[#2]{#2.}%
	}{%
		\oldparagraph[#1]{#2.}%
	}%
}

\ifsiam{
	\newlength\subsectionhspace
	\setlength\subsectionhspace{0pt plus 1em}
	
	\let\oldsubsection\subsection
	\renewcommand\subsection{%
	    \@ifstar{\mysubsectionStar}{\mysubsectionNoStar}%
	}
	\newcommand\mysubsectionStar[1]{%
		\oldsubsection*{#1.\hspace{\subsectionhspace}}%
	}
	\newcommand\mysubsectionNoStar[2][]{%
		\ifthenelse{\equal{#1}{}}{%
			\oldsubsection[#2]{#2.\hspace{\subsectionhspace}}%
		}{%
			\oldsubsection[#1]{#2.\hspace{\subsectionhspace}}%
		}%
	}

}{}

\colorlet{symmetriccolor}{black!10}
\colorlet{grayedout}{black!50}

\tikzset{
	inner node/.style={
		circle,draw,inner sep=1pt,
		minimum size=1.7em,
		fill=backgroundshading!50,
	},
	leaf node/.style={
		draw,rectangle,minimum size=1.25em,inner sep=1pt,
		fill=backgroundshading!75,
	},
	inner node concise/.style={
		circle,draw,fill,minimum size=.8ex,inner sep=0pt,
	},
	leaf node concise/.style={
		draw,fill,rectangle,minimum size=.5ex,inner sep=0pt,
	},
	leaf node empty/.style={
		draw=none,fill=none,rectangle,minimum height=.5ex,minimum width=.15ex,
	},
	partitioning node/.style={
		draw,inner sep=1pt,minimum height=1.5em,minimum width=1.5em,
		rectangle,rounded corners=2pt,
		fill=backgroundshading!50,
	},
	graphs/comparison tree detailed/.style={
		extended binary tree layout,
		nodes={inner node},
		leaf/.style={leaf node},
		math nodes,
		level distance=0ex, sibling distance=0ex,
		sibling sep=.6em, level sep=.4em,
	},
	graphs/comparison tree detailed compact/.style={
		extended binary tree layout,
		nodes={inner node,minimum size=1.05em,inner sep=0pt},
		leaf/.style={leaf node,minimum size=.6em,font=\scriptsize},
		math nodes,
		level distance=0ex, sibling distance=0ex,
		sibling sep=.6em, level sep=.4em,
	},
	graphs/recursion tree detailed/.style={
		tree layout,
		significant sep=.5em,
		missing nodes get space,
		nodes={partitioning node},
		leaf/.style={partitioning node},
		math nodes,
		level distance=0ex, sibling distance=0ex,
		sibling sep=.8em, level sep=.8em,
	},
	graphs/ternary comparison tree detailed/.style={
		tree layout,
		significant sep=.5em,
		missing nodes get space,
		nodes={inner node,minimum size=1.8em,},
		leaf/.style={leaf node},
		math nodes,
		edge quotes={
			pos=.33,
			draw,circle,fill=backgroundshading,
			inner sep=0pt,
			minimum size=8pt,
			font=\tiny,
			anchor=center,
		},
		level distance=0ex, sibling distance=0ex,
		sibling sep=1.1em, level sep=1.3em,
	},
	graphs/comparison tree concise/.style={
		binary tree layout,
		significant sep=0ex,
		nodes={inner node concise},
		empty nodes,
		leaf/.style={leaf node empty},
		level distance=0ex, sibling distance=0ex,
		sibling sep=.8ex, level sep=.4ex,
	},
}

\def\mydots{\xleaders\hbox to.5em{\hfill.\hfill}\hfill}

\makeatletter

\newlength\tmpLenNotations
\newenvironment{notations}[1][10em]{%
	\small
	\newcommand\notationentry[1]{%
		\settowidth\tmpLenNotations{##1}%
		\ifthenelse{\lengthtest{\tmpLenNotations > \labelwidth}}{%
			\parbox[b]{\labelwidth}{%
				\makebox[0pt][l]{##1}\\%
			}%
		}{%
			\mbox{##1}%
		}%
		\mydots\relax%
	}%
	\begin{list}{}{%
		\setlength\labelsep{0em}%
		\setlength\labelwidth{#1}%
		\setlength\leftmargin{\labelwidth+\labelsep+1em}%
	}
		\newcommand\notation[1]{\item[{##1}]}%
	\raggedright
}{%
	\end{list}
}

\let\oldthebibliography\thebibliography
\renewcommand\thebibliography[1]{%
	\oldthebibliography{#1}%
	\pdfbookmark[1]{References}{}%
	\markright{References}
	\let\path\nolinkurl%
}

\newsavebox\citehrefbox

\newsavebox\tabletmp%
\newlength\tmpheight%
\newlength\tmpdepth%

\let\epsilon\varepsilon

\ifsiam{

}{}

\counterwithout{equation}{section}

\newcommand{\Exp}[1]{\mathbb{E}\left[{#1}\right]}

\newcommand{\rfact}[2]{{#1}^{\overline{#2}}}

\newcommand{\poly}[1]{\mathopen{\langle\!\langle} {#1}\mathclose{\rangle\!\rangle}}

\newcommand\extendedversion{extended online version\xspace}

\makeatletter
\wref@putname{sec}{\S\!}
\wref@putname{fig}{Fig.\!}
\wref@putname{tab}{Tab.\!}
\wref@putname{thm}{Thm.\!}
\wref@putname{lem}{Lem.\!}

\usepackage[noend]{algpseudocode}

\ifsiam{
	\def\myemailsep{}
}{
	\def\myemailsep{\\}
}
\author{Conrado Martínez\thanks{%
		Department of Computer Science,
		Universitat Politècnica de Catalunya, Spain.
		\protect\myemailsep
		Email: \texttt{conrado\,@\,cs.upc.edu}%
        } 
        \and Markus Nebel\thanks{%
        Faculty of Technology,
        Technische Universität Bielefeld, Germany.
        \protect\myemailsep
        Email: \texttt{nebel\,@\,techfak.uni-bielefeld.de}
        } 
        \and Sebastian Wild\thanks{%
        David R.\ Cheriton School of Computer Science,
        University of Waterloo, Canada.
        \protect\myemailsep
		Email: \texttt{wild\,@\,uwaterloo.ca}%
	    }
}
\def\mytitle{Sesquickselect: One and a half pivots for cache-efficient selection}
\def\myfunding{%
	This work has been partially supported by funds from the 
	Spanish Ministery of Economy, Industry and Competitiviness (MINECO) 
	and the European Union (FEDER) under grant GRAMM (TIN2017-86727-C2-1-R), 
	and by funds from the Catalan Government (AGAUR) under grant 2017\,SGR 786.
	The last author is supported by the 
	Natural Sciences and Engineering Research Council of Canada 
	and the Canada Research Chairs Programme.
}
\ifsubmission{
	\title{%
			~\\[-2\baselineskip]%
			\mytitle%
		\thanks{\myfunding}%
	}
}{}
\ifarxiv{
	\title{\mytitle\thanks{\myfunding}}
}{}
\ifsiam{
	\title{\mytitle\thanks{\strut\myfunding}}
}{}

\ifsiam{
	\fancyfoot[R]{\scriptsize{Copyright \textcopyright\ 2019\\
	 Copyright for this paper is retained by authors.}}
}{}

\date{\small\today}
\graphicspath{{figs/}}

\begin{document}

\maketitle

\begin{abstract}
\ifsiam{}{%
	~\\\noindent\textbf{\textsf{Abstract:}\;\;}
}%
Because of unmatched improvements in CPU performance,
memory transfers have become a bottleneck of program execution.
As discovered in recent years, this also affects sorting in internal memory.
Since partitioning around several pivots reduces overall memory transfers,
we have seen renewed interest in multiway Quicksort.
Here, we analyze in how far multiway partitioning helps in Quickselect.

We compute the expected number of comparisons and scanned elements 
(approximating memory transfers) for a generic class of (non-adaptive) multiway Quickselect
and show that three or more pivots are not helpful, but two pivots are.
Moreover, we consider ``adaptive'' variants which choose partitioning and 
pivot-selection methods in each recursive step from a finite set of alternatives 
depending on the current (relative) sought rank.
We show that ``Sesquickselect'', a new Quickselect variant that uses either one or two pivots,
makes better use of small samples \wrt memory transfers than other Quickselect variants.
\end{abstract}

\section{Introduction}
\label{sec:intro}

We consider the selection problem: finding the $m$th
smallest element within an unsorted array of $n$ distinct elements.
Quickselect~\cite{Hoare1961} is the earliest (published) algorithm
for general selection that runs in linear time (in expectation),
and it forms the basis of practical implementations and more advanced algorithms.
From a theoretical perspective, this problem might be considered solved:
The Floyd-Rivest algorithm \cite{FloydRivest1975,Kiwiel2005}
uses $n+\min\{m,n-m\} + o(n)$ comparisons
in expectation, 
and this is optimal up to lower order terms~\cite{CuntoMunro1989}. 
The Floyd-Rivest algorithm is a variant of Quickselect
that uses a large
random sample of $\Theta(n^{2/3}\log^{1/3} n)$ elements
from which it (recursively) selects two pivots $P_1$ and $P_2$, $P_1<P_2$,
so that their ranks surround $m$ with high probability.
Partitioning the input into the elements
$<P_1$, between $P_1$ and $P_2$, and $>P_2$, respectively,
yields a subproblem of size $o(n)$ with high probability.

Standard libraries do not use the asymptotically optimal 
algorithms~\cite{Valois2000,Alexandrescu2017}, presumably
because for moderate-size inputs, lower order terms and their large hidden constants 
are not negligible, making variants with less overhead desirable.
For example, the GNU implementation of the C++ STL uses introspective 
median-of-3 Quickselect
for \texttt{std:\!:nth\_element}.%
\footnote{
	\ifsiam{\strut}{}%
	The code can be browsed online: 
	\url{https://gcc.gnu.org/onlinedocs/gcc-8.1.0/libstdc++/api/a00527_source.html\#l04748}.%
}
Introspective sorting was suggested by Musser~\cite{Musser1997}
and refined by Valois~\cite{Valois2000}.

Given the similarity of Quicksort and Quickselect, it is natural and tempting to employ the same
optimizations for selection that work well for sorting. 
Indeed, this is exactly what is done in the GNU STL;
the Quickselect implementation uses 
the same partitioning method,
the same pivot rule (median of three elements),
the same protection mechanism against bad-case inputs
(a recursion-depth limit of $2\lceil\lg(n+1)\rceil$ with a $\Theta(n\log n)$ worst-case method
based on heapsort), and 
the same base case for the recursion (Insertionsort)
as in the Quicksort implementation.
On second thought, some of these design decisions are quite questionable in the context of selection.
First, a linear worst case can be achieved instead of the $\Theta(n \log n)$
one~\cite{BlumFloydPrattRivestTarjan1973,Valois2000}.
Second, it is known that choosing the pivot \emph{adaptively,} 
\ie, depending on the value of $m$ 
(mimicking the asymptotically optimal Floyd-Rivest algorithm!)
improves the average costs by a significant factor
even for small sample sizes~\cite{MartinezPanarioViola2010}.

Finally, in light of the recent success of multi-pivot Quicksort%
~\cite{AumuellerDietzfelbingerKlaue2016,Kushagra2014,NebelWildMartinez2016,Wild2016},
a question programmers will face is whether and how 
multiway partitioning should also be used for Quickselect.
Using YBB-partitioning%
\footnote{
	\ifsiam{\strut}{}%
	Named after its inventors 
	Vladimir \underline Yaroslavskiy, 
	Jon L. \underline Bentley, and 
	Joshua \underline Bloch.
}%
~-- 
the dual-pivot method used in \texttt{Arrays.sort} of Oracle's Java runtime library%
~\cite{Yaroslavskiy2009,WildNebelNeininger2015}~-- 
was shown to be of \emph{no} advantage for Quickselect \wrt 
the number of comparisons (averaging over all possible ranks
to be selected)~\cite{WildNebelMahmoud2016}.
YBB partitioning can reduce the comparison count in sorting,
but the main advantage of multiway partitioning 
lies in saving memory transfers, and indeed, the latter \emph{is} 
improved using dual-pivot Quickselect (see \wref{sec:random-ranks}).
The purpose of this article is thus 
a comprehensive assessment of the 
potential of multiway partitioning in Quickselect.

To this end, we present an average-case analysis of 
both classical cost measures and memory transfers.
The latter is formalized as the number of \emph{scanned elements}:
the accumulated range scanned by all index/pointer variables
used in the partitioning strategy.
This has been shown to be a good indicator for the number of 
cache misses that occur during partitioning~\cite{NebelWildMartinez2016,AumuellerDietzfelbingerKlaue2016}.

Our focus is on low-overhead algorithms suitable for library implementations,
and hence on small fixed-size samples.
Taking inspiration from Floyd-Rivest, we propose ``Sesquickselect'',%
\footnote{
	After the Latin prefix \textit{sesqui-} meaning ``one and a half''.
}
an adaptive Quickselect variant that uses either one or two pivots
from a sample of $k$ elements, and we give strong evidence
for its optimality \wrt scanned elements subject to a given sample size.

\paragraph{Overview and Method}
We first consider multiway partitioning and pivot sampling in full generality
(partitioning into any constant number $s\ge 2$ of segments 
while choosing pivots from samples of any constant size~$k$),
under the assumption that a uniformly chosen random rank is searched.
These so-called ``grand averages''~\cite{MahmoudModarresSmythe1995} 
can be computed using a distributional master theorem~\cite{Wild2016}
derived from Roura's continuous master theorem~\cite{Roura2001}.
We can conclude that
\textsl{more than three segments 
are indeed \textit{not} helpful in Quickselect.} 
This matches the intuition that a large $s$ ``should'' not help
since all but one segment will be discarded for good, 
making further subdivisions superfluous.
The precise argument requires some care, though.

Unfortunately, the grand-average analysis does not extend to
adaptive methods like Sesquickselect.
For the second part, we hence consider selecting the $\alpha$-quantile in a large array
for a fixed $\alpha\in(0,1)$ (extending techniques from~\cite{MartinezPanarioViola2010}).
We give an elementary proof for the correctness of a resulting integral equation
for the leading-term coefficient as a function of $\alpha$ 
(under reasonable assumptions fulfilled for our applications)
that appears to be novel.

The setting with two parameters makes computations appreciably more challenging.
For Quickselect with YBB partitioning (without pivot sampling) and 
Sesquickselect with $k=2$ we  solve the integral equations analytically,
and we obtain precise numerical solutions for more general cases.
From these, we can derive promising candidates of
cache-optimal Quickselect variants for all practical sample sizes.

\paragraph{Outline}
We give an overview of previous work in the remainder of this section.
\wref{sec:prelim} introduces notation and preliminaries.
In \wref{sec:distributional-recurrence}, we state a general distributional
recurrence of costs.
\wref{sec:random-ranks} discusses the analysis for random ranks. 
We then switch to fixed ranks and derive the integral equation in \wref{sec:linear-ranks-convergence}.
We solve it for Quickselect with YBB-partitioning 
(\wref{sec:linear-ranks-ybb}) and for the novel ``Sesquickselect'' algorithm (\wref{sec:sesquickselect}). 
Our paper concludes with a discussion of
our findings (\wref{sec:conclusion}).

\ifsiam{%
	There is an \extendedversion of this paper available as arxiv preprint
	that contains missing proofs and computations.%
}{%
	The appendix contains a comprehensive list of notation, as well as
	some technical proofs and details of the computations.
}

\subsection{Previous Work}
\label{sec:previous-work}

The first published analysis of (classic) Quickselect by Knuth
served as one of two illustrating examples in an invited address 
at the IFIP Congress 1971, with the goal to advertise the emerging area of 
analysis of algorithms~\cite{Knuth1971,Knuth2000}.
The expected number of comparisons in classic Quickselect is
\(
		\E{C_{n,m}} 
	\wwrel= 
		2\bigl((n+1)\harm n - (n+3-m)\harm{n+1-m} - (m+2) \harm m + n+3\bigr),
\)
where $\harm n = \sum_{i=1}^n\!\frac1i$.

An asymptotic approximation for selecting 
the $\alpha$-quantile, $\alpha\in(0,1)$ fixed, 
follows with $m=\alpha n$:
\(
		\E{C_{n,\alpha n}} 
	\wwrel= 
		2(h(\alpha) + 1)\cdot n \bin- 8 \ln n \wbin\pm \Oh(1)
\),
		$n\to\infty$,
where $h(x) = -x\ln x - (1-x)\ln (1-x)$;
(here we set $0 \ln 0 \ce 0$).

Quickselect has been extensively studied.
The variance is quadratic and precisely known~\cite{Paulsen1997,Kirschenhofer1998};
large deviations from the mean are very unlikely~\cite{Devroye1984,Gruebel1998}.
Stochastic limits laws have been established for the random costs divided by $n$ 
for random ranks~\cite{MahmoudModarresSmythe1995},
small ranks~\cite{HwangTsai2002} and fixed quantiles~\cite{GruebelRoesler1996}.

Like for Quicksort, better pivot selection methods are important in Quickselect.
The widely used median-of-three version was analyzed in~\cite{AndersonBrown1992} 
and~\cite{KirschenhoferProdingerMartinez1997},
and the generalization of using the median of any fixed size sample
(``median-of-$k$'') in~\cite{MartinezRoura2001} (expectation and variance for random ranks) and 
in~\cite{Gruebel1999} (for fixed ranks); 
refined limit laws for the case where $k$ grows polynomially with $n$ were recently 
derived in~\cite{SulzbachNeiningerDrmota2014}.
Choosing the order statistic of the sample depending on $m/n$ was studied
in~\cite{MartinezPanarioViola2010} (\wrt expectation)
and~\cite{KnofRoesler2012} (limit laws, including growing~$k$).

If the objects to select from are strings, symbol comparisons rather than key comparisons
are the measure of interest; Quickselect has linear expected cost
also in this model~\cite{ClementFillThiValee2015}.
Quickselect with YBB partitioning was analyzed in~\cite{WildNebelMahmoud2016} and
Krenn~\cite{Krenn2017} considered the comparison-optimal
dual-pivot partitioning method of Aumüller and Dietzfelbinger~\cite{AumullerDietzfelbinger2015}.

\section{Preliminaries}
\label{sec:prelim}

\ifsiam{
	We start with some notation.%
}{%
We introduce important notation here;
see \wref{app:notations} for a comprehensive list.%
}
$\Oh$-terms are bounds on the absolute value; we write $g = f \pm \Oh(e)$ to emphasize 
this. $f\sim g$ means $f / g \to 1$.
Vector's are written in boldface, \eg, $\vect x=(x_1,x_2,x_3)$, 
and operations are understood componentwise: $\vect x+1 = (x_1+1,x_2+1,x_3+1)$. 
We use the notation $x^{\overline n}$ and $x^{\underline n}$ 
of~\cite{ConcreteMathematics} for rising resp.\ falling (factorial) powers.
(The former is also known as Pochhammer function).

We use capital letters for random variables and $\ESymbol$ for expectation.
$X\eqdist Y$ denotes equality in distribution.
$\uniform[1..n]$ is the discrete uniform distribution over $[1..n]$, 
$\uniform(0,1)$ the continuous uniform distribution over $(0,1)$.
We next recall the beta distribution and some of its properties.
It will play a pivotal role in our analysis.

\subsection{The beta distribution and its relatives}

The \emph{beta distribution} has two parameters $\alpha,\beta\in\R_{>0}$ 
and is written as $\betadist(\alpha,\beta)$.
If $X\eqdist\betadist(\alpha,\beta)$, we have $X \in (0,1)$ and $X$ has the density 
\begin{align*}
		f(x)
	&\wrel=
		\frac{x^{\alpha-1}(1-x)^{\beta-1} }{\BetaFun(\alpha,\beta)}
		,\qquad x\in(0,1),
\end{align*}
where $\BetaFun(\alpha,\beta) = \Gamma(\alpha)\Gamma(\beta) / \Gamma(\alpha+\beta)$
is the beta function.
The reason why the beta arise in our analysis is
its connection to order statistics:
If we assume the input consists of $n$ \iid (independent and identically distributed) 
$\uniform(0,1)$ random variables, the $\ell$th smallest element of a sample of $k$ elements
has a $\betadist(\ell,k+1-\ell)$ distribution.

For multi-pivot methods, we will encounter the Dirichlet distribution $\dirichlet(\vect\alpha)$,
which is the multivariate version of the beta distribution.
For $X\eqdist \dirichlet(\vect\alpha)$ with $\vect\alpha\in\R_{>0}^d$,
we use the convention that $X_d = 1 - X_1- \cdots- X_{d-1}$ and specify $\vect X$ as a $d$-dimensional vector.
Then $\vect X$ has the density
$f(\vect x) = \vect x^{\vect\alpha-1} / \BetaFun(\vect\alpha)$,
where $\BetaFun(\vect\alpha) = \Gamma(\alpha_1)\cdots\Gamma(\alpha_d) / \Gamma(\alpha_1+\cdots\alpha_d)$
is the $d$-dimensional beta function.

For subproblem sizes, we will furthermore find the Dirichlet-multinomial distribution 
$\dirichletMultinomial(n,\vect\alpha) \eqdist \multinomial(n,\dirichlet(\vect\alpha))$,
which is a mixed multinomial distribution with a Dirichlet-distributed parameter.
For the 2d case, the distribution is called beta binomial distribution,
written as $\betaBinomial(n,\alpha,\beta)$.

Since the binomial distribution is sharply concentrated, one can
use Chernoff bounds to show that $\betaBinomial(n,\alpha,\beta)/n$ converges to $\betadist(\alpha,\beta)$
in a specific sense.
We can obtain stronger error bounds by directly comparing the PDFs,
and the argument generalizes to higher dimensions.
The two-dimensional case appears in \cite[Lemma~2.38]{Wild2016};
here we extend it to general (fixed) $s\ge 2$.
We following the notation used there, in particular we write 
$\total{\vect x} = \sum_{i=1}^s x_i$.

\begin{lemma}[Local Limit Law \ifsiam{DirMult}{for Dirichlet-multinomial}]
\label{lem:local-limit-law-dirichlet-multinomial}
	\oneline{Let \((\ui{\vect I}n)_{n\in\N_{\ge1}}\)} 
	be a family of random variables with Dirichlet-multinomial distribution,
	\(\ui{\vect I}n \eqdist \dirichletMultinomial(n, \vect \alpha)\) where 
	\(\vect \alpha\in(\{1\}\cup\R_{\ge2})^s\) is fixed, 
	and let \(f_D(\vect z)\) be the density of the \(\dirichlet(\vect\alpha)\) distribution.
	Then we have uniformly in \(\vect z\in(0,1)^s\) (with $\total{\vect z} = 1$) that 
	\begin{align*}
				n^{s-1} \cdot \Prob[\big]{ \ui{\vect I}n = \lfloor \vect z(n+1)\rfloor } 
			\wwrel= 
				f_D(\vect z) \bin\pm \Oh(n^{-1}),
	\end{align*}
	as $n\to\infty$.
	That is, \(\ui {\vect I}n/n\) converges to \(\dirichlet(\vect\alpha)\) in distribution, 
	and the probability weights converge uniformly to the limiting density at rate \(\Oh(n^{-1})\).
\end{lemma}
\ifsiam{%
	The proof is given in the \extendedversion.
}{%
	The proof is given in \wref{app:local-limit-beta}.
}

\begin{remark}
	Since $f_D$ is a polynomial in $\vect z$, it is in particular bounded and Lipschitz-continuous
	in the closed domain $\vect z \in [0,1]^s$ with $\total{\vect z}=1$.
	Hence, the local limit law also holds for the random variables $\ui{\vect I}n + c$ for any constant $c$.
	We use this for subproblem sizes, which are of this form: $J_r = I_r + t_r$.
\end{remark}

\subsection{Hölder-Continuity}

A function $f:I\to \R$ defined on a bounded interval $I$ is 
called Hölder-continuous with exponent $h\in(0,1]$
when 
\[
	\exists C\;
	\forall x,y\in I\wrel:
		\bigl| f(x) - f(y) \bigr|
		\wrel\le 
		C |x-y|^h.
\]
Hölder-continuity is a form of smoothness of functions
that is stricter than (uniform) continuity, but slightly more liberal
than Lipschitz-continuity (which corresponds to $h=1$).
It provides a useful requirement in some of our theorems;
$f:[0,1]\to\R$ with $f(z) = z \ln(1/z)$ is a stereotypical function
that is Hölder-continuous (for any $h\in(0,1)$), but not Lipschitz.
We will also need Hölder-continuity for functions from $\R^n$ to $\R$;
we extend the definition by requiring
\(
		\bigl| f(\vect x) - f(\vect y) \bigr|
	\wrel\le 
		C \|\vect x - \vect y\|^h
\),
\ie, using the Euclidian norm.

The most useful consequence of Hölder-continuity is given by the following lemma:
an error bound on the difference between an integral and the Riemann sum.

\begin{lemma}
\label{lem:hölder-intergral-bound}
	Let $f:[0,1]^d \to \R$ be Hölder-continuous (\wrt $\|\cdot\|_2$) with
	exponent $h$.
	Then
	\begin{align*}
			\int_{\vect x\in [0,1]^d} f(\vect x) \, d\vect x
		&\wwrel=
			\frac1{n^d} \!\! \sum_{\vect i \in [0..n-1]^d} \!\!\! f(\vect i / n)
			\wwbin\pm
			\Oh(n^{-h}),
	\end{align*}
	as $n\to\infty$.
\end{lemma}
\ifsiam{%
	The proof is a simple computation; it is given
	in the \extendedversion.
}{%
	The proof is given in \wref{app:hölder}.
}

We considered only the unit interval resp.\ unit hypercube
as the domain of functions rather than a product of general compact intervals,
but this is no restriction:
Hölder-continuity (on bounded domains) is preserved by addition,
subtraction, multiplication and composition
(see, \eg, \cite[\S\,4.6]{Sohrab2014}).
Since any linear function is Lipschitz, the above result holds for 
Hölder-continuous functions $f:[a_1,b_1]\times\cdots\times[a_d,b_d]\to \R$.

If our functions are defined on a bounded domain,
Lipschitz-continuity implies Hölder-continuity
and Hölder-continuity with exponent $h$ implies
Hölder-continuity with exponent $h' < h$.
A real-valued function is Lipschitz if its derivative is bounded resp.\ 
if all its partial derivatives are bounded (in the multivariate case).
The latter follows form the mean-value theorem (in several variables)
and the Cauchy-Schwartz inequality.

\subsection{The Distributional Master Theorem}
\label{sec:DMT}

To solve the recurrences in \wref{sec:random-ranks},
we use the ``distributional master theorem'' (DMT)%
~\cite[\href{https://www.wild-inter.net/publications/html/wild-2016.pdf.html\#pf8a}{Thm.~2.76}]{Wild2016}, 
reproduced below for convenience.
It is based on Roura's continuous master theorem~\cite{Roura2001}, 
but reformulated in terms of distributional recurrences in an attempt to give 
the technical conditions and occurring constants in Roura's original formulation 
a more intuitive, stochastic interpretation.
We start with a bit of motivation for the latter.

The DMT is targeted at divide-and-conquer recurrences where the recursive parts
have a \emph{random} size like in Quickselect.
Because of the random subproblem sizes, a traditional recurrence for expected costs
has to sum over all possible subproblem sizes, weighted appropriately.
That way, the direct correspondence between the recurrence and the algorithmic process 
is lost.
A \emph{distributional recurrence} avoids this.
It describes the full distribution of costs, where
the cost for larger problem sizes is described by a 
``toll term'' (the partitioning costs in Quickselect) 
plus the contributions of recursive calls.

Such a distributional formulation requires 
the toll costs and subproblem sizes to be stochastically 
independent of the recursive costs when conditioned on the subproblem sizes.
In typical applications, this is fulfilled when the studied algorithm 
guarantees that the subproblems on which it calls itself recursively
are of the same nature as the original problem.
Such a form of randomness preservation is also required for the analysis
using traditional recurrences.

The DMT allows us to compute an asymptotic approximation of the expected costs
directly from the distributional recurrence.
Intuitively speaking, it is applicable whenever the \emph{relative} subproblem sizes
of recursive applications converge to a (non-degenerate) limit distribution as $n\to\infty$
(in a suitable sense; see \weqref{eq:DMTwc-condition} below).
The local limit law provided by \wref{lem:local-limit-law-dirichlet-multinomial}
gives exactly such a limit distribution.

\savebox\citehrefbox{%
\bfseries\cite[\href{https://www.wild-inter.net/publications/html/wild-2016.pdf.html\#pf8a}{Thm.~2.76}]{Wild2016}%
}%
\begin{theorem}[DMT \usebox\citehrefbox]
\label{thm:DMTwc}
	Let \((C_n)_{n\in\N_0}\) be a family of random variables that
	satisfies the distributional recurrence
	\begin{align}
	\label{eq:DMTwc-distributional-recurrence}
			C_n
		\wwrel\eqdist
			T_n \bin+ \sum_{r=1}^s \ui{A_r}n \cdot C_{\ui{J_r}n}^{(r)},
			\qquad (n \ge n_0),
	\end{align}
	where the families \((\ui{C_n}1)_{n\in\N},\ldots,(\ui{C_n}s)_{n\in\N}\) are independent copies of
	\((C_n)_{n\in\N}\), which are also independent of 
	\((\ui{J_1}n,\ldots,\ui{J_s}n)\in\{0,\ldots,n-1\}^s\),
	\((\ui{A_1}n,\ldots,\ui{A_s}n) \in \R_{\ge0}^s\)
	and \(T_n\).
	Define \(\ui{Z_r}n = \ui{J_r}n / n\), $=1,\ldots,s$, and assume that they 
	fulfill uniformly for \(z\in(0,1)\)
	\begin{align}
	\label{eq:DMTwc-condition}
			n \cdot \Prob[\big]{ \ui{Z_r}n \in (z-\tfrac1n,z] }
		&\wwrel= 
			f_{Z_r^*}(z) \bin\pm \Oh(n^{-\delta}),
	\end{align}
	as $n\to\infty$ for a constant \(\delta>0\) and a 
	Hölder-continuous function \(f_{Z_r^*} : [0,1] \to \R\).
	Then \(f_{Z_r^*}\) is the density of a random variable \(Z_r^*\) and
	\(\ui{Z_r}n \convD Z_r^*\).

	Let further 
	\begin{align}
	\label{eq:DMTwc-condition-coeffs}
			\E[\big]{ \ui{A_r}n \given \ui{Z_r}n \in (z-\tfrac1n,z] }
		&\wwrel=
			a_r(z) \wbin\pm \Oh(n^{-\delta}),
	\end{align}
	as $n\to\infty$ for a function \(a_r : [0,1] \to \R\) and
	require that \(f_{Z_r^*}(z)\cdot a_r(z)\) is also Hölder continuous on~\([0,1]\).
	Moreover, assume \(\E{T_n} \sim K n^\alpha \log^\beta(n)\), as \(n\to\infty\), 
	for constants \(K\ne 0\), \(\alpha\ge0\) and \(\beta>-1\).
	Then, with \(H = 1 - \sum_{r=1}^s\E{(Z_r^*)^\alpha a_r(Z_r^*)}\), we have the following cases.
	\begin{enumerate}[itemsep=0ex]
	\item If \(H > 0\), then \(\displaystyle \E{C_n} \sim \frac{\E{T_n}}{H}\).
			\(\vphantom{\displaystyle\sum_{i}^{i}}\)
	\item \label{case:DMTwc-H0} 
		If \(H = 0\), then 
		$\displaystyle
		\E{C_n} \sim \frac{\E{T_n} \ln n}{\tilde H}$ with 
		$\displaystyle \tilde H = -(\beta+1)\sum_{r=1}^s 
			\E{(Z_r^*)^\alpha a_r(Z_r^*) \ln(Z_r^*)}$.
	\item \label{case:DMTwc-theta-nc}
		If \(H < 0\), then \(\E{C_n} = \Oh(n^c)\) for the
		\(c\in\R\) with \(\displaystyle\sum_{r=1}^s\E{(Z_r^*)^c a_r(Z_r^*)} = 1\).
	\end{enumerate}
\qed\end{theorem}

\subsection{Adaptive Quickselect}
\label{sec:adaptive-quickselect}

We consider the following generic family of Quickselect variants:
In each step, we partition the input into $s\ge 2$ segments, 
choosing the $s-1$ pivots $P_1,\ldots,P_{s-1}$
as order statistics from a random sample of the input.
The pivot-selection process is described by two parameters:
the sample size $k$ and the quantiles vector $\vect \tau = (\tau_1,\ldots,\tau_s) \in [0,1]^s$.
The $\ell$th pivot, $\ell=1,\ldots,s-1$, is the $(\tau_1+\cdots+\tau_\ell)$-quantile of the sample.
Alternatively, we use the vector $\vect t = (k+1)\vect \tau - 1 \in \N_{\ge0}^s$ 
to specify how many
elements are \emph{omitted} between the pivots.
As an example, median-of-$3$ (for classic $s=2$) corresponds to
$k=3$, $\vect\tau=(\frac12,\frac12)$ or $\vect t = (1,1)$.
Choosing the two largest elements in a sample of $5$ corresponds to $k=5$, 
$\vect\tau=(\frac23,\frac16,\frac16)$ or $\vect t=(3,0,0)$.
Note that $\tau_\ell$ is the \emph{expected fraction} 
of elements in the $\ell$th segment, $\ell=1,\ldots,s$.
See \wref{fig:notations} for another example.

We assume the partitioning algorithm is an instance of the generic one-pass partitioning scheme
analyzed in~\cite{Wild2016} which unifies practically relevant methods.
(A similar such scheme is considered in~\cite{AumuellerDietzfelbingerKlaue2016}).
The details of the partitioning method are mostly irrelevant for our present discussion,
and we refer the reader to 
\cite[\href{https://www.wild-inter.net/publications/html/wild-2016.pdf.html\#pfaa}{\S4.3}]{Wild2016} for details;
important here is that partitioning preserves randomness for recursive calls
and that the expected partitioning costs are $a n \pm \Oh(1)$
for a known constant $a$ (depending only on the partitioning method and $\vect t$).

We consider \emph{adaptive} Quickselect variants, which are formally given by 
specifying the partitioning method and parameters $s$ and $\vect t$
as a \emph{function} of $\alpha = \frac mn$.
We assume a fixed, finite portfolio of methods to choose from and the choice 
consists in finding the interval containing $\alpha$ in a finite collection
$I_1,\ldots,I_d$ of intervals (with $I_1\cup\cdots\cup I_d = [0,1]$).
We treat the parameters as functions of $\alpha$ with the meaning
$s(\alpha) = s_{v}$ for the $v\in[d]$ with $\alpha \in I_v$,
and similarly for $\vect t(\alpha) = \ui{\vect t}v$
and $a_{\mathcal F}(\alpha) = \ui{a_{\mathcal F}}v$
(introduced in the next section).%
As a specific example for an adaptive method, consider Sesquickselect
with a sample of size $k=2$.
There, we can choose $s$ and $\vect t$ as follows:
\begin{align*}
		s(\alpha)
	&\wwrel=
		\begin{cases}
			1, & \text{if } \alpha < 0.266; \\
			2, & \text{if } 0.266 \le \alpha \le 0.734; \\
			1, & \text{if } \alpha > 0.734;
		\end{cases}
\\
		\vect t(\alpha)
	&\wwrel=
		\begin{cases}
			(0,1),   & \text{if } \alpha < 0.266; \\
			(0,0,0), & \text{if } 0.266 \le \alpha \le 0.734; \\
			(1,0),   & \text{if } \alpha > 0.734.
		\end{cases}
\end{align*}
Other combinations are of course possible, but
we will show in \wref{sec:sesquickselect} 
that this is indeed a good choice.

\subsection{Cost Measures and Notation}

We consider several measures of cost for Quickselect:
$\mathcal C$, the number of key comparisons,
$\mathcal{SE}$, the number of scanned elements (total distance traveled by pointers / scanning indices),
and $\mathcal{WA}$, the number of write accesses to the array.%
\footnote{%
	\ifsiam{\strut}{}%
	Write accesses are more appropriate than counting swaps for methods that 
	move several elements at a time.%
}
The analysis can mostly remain agnostic to this in which case
we use $\mathcal F$ as placeholder for any of the above.
We use the following naming conventions for the quantities arising in our analysis:

\vspace{.5\baselineskip}

\begin{itemize}[nosep]
	\item $F_{n,m}$ for the random costs to select the $m$th smallest out of $n$ elements.
	\item $F_{n,M_n}$ is the random cost to select a (uniform) \emph{random rank} 
		$M_n \eqdist \uniform[1..n]$ from $n$ elements.
	\item $f(\alpha)$ is the leading-term coefficients of $\E{F_{n,m}}$ for $n\to\infty$ and $m/n\to\alpha$.
	\item $\bar f$ is the leading-term coefficient of the grand average: 
		$\bar f = \lim_{n\to\infty} \E{F_{n,M_n}}/{n}$.
	\item $A_{\mathcal F}(n,m)$ are the random costs of the first partitioning round;
		they indirectly depend on $m$ for adaptive methods.
	\item $a_{\mathcal F}(\alpha)$ is the leading-term coefficient of partitioning costs
		for $n\to\infty$ and $m/n\to\alpha$.
\end{itemize}

\vspace{.5\baselineskip}

\section{Distributional Recurrence}
\label{sec:distributional-recurrence}

We will focus on the expected costs, $\E{F_{n,m}}$,
but a concise description can be given for the full distribution. 
The family of random variables $(F_{n,m})_{n\in\N,m\in[n]}$ fulfills 
the following distributional recurrence
(notation explained below)
\begin{align}
\label{eq:dist-rec-general}
		F_{n,m}
	&\wrel\eqdist
		A_{\mathcal F}(n,m) \bin+
		\sum_{\ell=1}^s \indicator{R_{\ell-1} < m < R_{\ell}} \ui{F_{J_\ell, \, m-R_{\ell-1}}}{\ell}
\end{align}
for $n\ge n_0$;
for small $n < n_0$ costs are given by some base-case method
that contributes only $\Oh(1)$ to overall costs.
In the general setting, we partition the input into $s\ge2$
segments around $s-1$ pivot elements.
We denote by $\ui{\vect J}n = (\ui{J_1}n,\ldots,\ui{J_s}n)$ the (vector of) resulting 
subproblem sizes (for recursive calls).
$\ui{R_1}n\le\cdots\le \ui{R_{s-1}}n$ are the (random) \emph{ranks} of the pivot elements;
we set $\ui{R_0}n = 0$ and $\ui{R_{s}}n = n+1$ to unify notation for the outermost segments.
As in \wref{eq:dist-rec-general}, we usually suppress the dependence on $n$ in our notation
for better legibility.
Note that pivot ranks 
and subproblem sizes %
are related via $J_\ell = R_\ell - R_{\ell- 1} -1$ for $\ell = 1,\ldots,s$
(cf.\ \wref{fig:notations}).
For $\ell=1,\ldots,s$, $(\ui{F_{n,m}}\ell)_{n\in\N,m\in[n]}$ are 
independent copies of $(F_{n,m})_{n\in\N,m\in[n]}$,
which are also independent of $\ui{\vect J}n$ 
(and hence $\ui{R_1}n,\ldots,\ui{R_{s-1}}n$) and $A_{\mathcal F}(n,m)$.

The distribution of the subproblem sizes~$\vect J$ is discussed in detail below
for the standard non-adaptive case (\wref{sec:distribution-subproblem-sizes}).
We remark that \weqref{eq:dist-rec-general} remains valid for \emph{adaptive} variants,
\ie, where the employed pivot sampling scheme and partitioning method are chosen 
in each step depending on $\alpha = \frac mn$, the relative rank of the sought element.
(The distributions of $\vect J$ and $\vect R$ are then functions of $n$ \emph{and} $m$.)

\section{Random ranks}
\label{sec:random-ranks}

In this section we consider $\E{F_{n,M_n}}$,
where $M_n \eqdist \uniform[1..n]$. 
This ``grand average''~\cite{MahmoudModarresSmythe1995} is a reasonable measure 
for a rough comparison of different selection methods,
and its analysis is feasible for a large class of algorithms.
Indeed,
an asymptotic approximation of the costs will follow from 
a distributional master theorem (DMT, \wref{thm:DMTwc}).
We consider only non-adaptive methods in this section,
\ie, the splitting probabilities do not depend on the rank of the sought element.

We will 
derive an asymptotic approximation for the grand average 
for a whole class of Quickselect variants 
that cover the above special cases as well as many further hypothetical versions.
Our partitioning method splits the input into $s\ge2$
segments using $s-1$ pivot elements and the pivot elements are selected
as order statistics from a fixed-size random sample of the input.

\begin{figure}[htbp]
	\plaincenter{
	\externalizedpicture{notations}
	\begin{tikzpicture}[scale=.35]
		\def\n{25}\def\npone{26}
		\def\Rone{4}
		\def\Rtwo{13}
		\def\Rthree{21}
		\def\Pone{5/25}
		\def\Ptwo{14.5/25}
		\def\Pthree{22/25}
		\def\ps{6pt}
		\begin{scope}
			\node at (\n/2,6.5) {\textbf{Rank-Based World}} ;
			\node[align=left,font=\scriptsize,anchor=east] at (\n+1,5) 
				{$s=4$\\$n=25$\\$\vect t = (1,2,2,1)$\\$k=9$} ;
			
			\begin{scope}[shift={(\n/5,3)},scale=.8]
				\node[anchor=east] at (-.5,.5) {sample} ;
				\fill[pattern=north east lines, pattern color=black!50] (0,0) rectangle ++(9,1);
				\foreach \i/\r in {1/1,4/2,7/3} {
					\fill[black!30] (\i,0) rectangle node[black,scale=.5] {$P_\r$} ++(1,1) ;
				}
				\foreach \i in {0,...,9} {
					\coordinate (sample\i) at (\i,0);
				}
				\draw (0,0) grid ++(9,1);
				\draw[thick] (0,0) rectangle ++(9,1);
				\begin{scope}[decoration={brace,aspect=.5,amplitude=1},shift={(0,.2)}]
					\draw[decorate] (0,1) -- node[above,scale=.6] {$t_1$} (1,1);
					\draw[decorate] (2,1) -- node[above,scale=.6] {$t_2$} (4,1);
					\draw[decorate] (5,1) -- node[above,scale=.6] {$t_3$} (7,1);
					\draw[decorate] (8,1) -- node[above,scale=.6] {$t_4$} (9,1);
				\end{scope}
			\end{scope}
			
			\begin{pgfonlayer}{background}
			\begin{scope}[fill=black!10,very thin,draw=black!50]
				\foreach \x in {%
						0,\n,
						\Rone,\Rtwo,\Rthree,%
						\Rone-1,\Rtwo-1,\Rthree-1%
				} {
					\draw[black!50,thin,dotted] (\x,0) -- ++ (0,-3);
				}
				\filldraw (sample0) -- (1-1,1)  -- ++ (1,0) -- (sample1);
				\filldraw (sample1) -- (\Rone-1,1)  -- ++ (3,0) -- (sample4);
				\filldraw (sample4) -- (\Rtwo-1,1)  -- ++ (3,0) -- (sample7);
				\filldraw (sample7) -- (\Rthree-1,1)  -- ++ (2,0) -- (sample9);
			\end{scope}
			\end{pgfonlayer}
			
			\foreach \i/\r/\t in {1/\Rone/2,2/\Rtwo/2,3/\Rthree/1} {
				\fill[black!30] (\r,0) rectangle ++(-1,1) ;
				\fill[pattern=north east lines, pattern color=black!50] (\r,0) rectangle ++(\t,1);
			}
			\foreach \i/\r/\t in {0/0/1,4/\npone/0} {
				\fill[black!20] (\r,0) rectangle ++(-1,1) ;
				\fill[pattern=north east lines, pattern color=black!50] (\r,0) rectangle ++(\t,1);
			}
			\draw (0,0) grid (\n,1);
			\draw[thick] (0,0) rectangle (\n,1);
			\foreach \i in {0,...,\npone} {
				\node at (\i-0.5,1.5) {\tiny \i} ;
			}
			\foreach \i/\r in {0/0,1/\Rone,2/\Rtwo,3/\Rthree,4/\npone} {
				\path (\r-.5,0) ++(0,-.75) node[scale=1] {\scriptsize 
					$
						\llap{\ifthenelse{\equal{\i}{0}}{$0={}$}{}}
						R_\i 
						\rlap{\ifthenelse{\equal{\i}{4}}{${}{=}n{+}1$}{}}
					$
				} ;
				\node[scale=.8] at (\r-.5,.5) {\scriptsize 
					$
						\llap{\ifthenelse{\equal{\i}{0}}{$0={}$}{}}
						P_\i 
						\rlap{\ifthenelse{\equal{\i}{4}}{${}=1$}{}}
					$
				} ;
			}

			\foreach \s/\e/\l in {1/\Rone-1/J_1,\Rone+1/\Rtwo-1/J_2,\Rtwo+1/\Rthree-1/J_3,\Rthree+1/\n/J_4} {
				\draw[decorate,decoration={brace,mirror,amplitude=4}] (\s-1,-2) 
					-- node[below=5pt,scale=.8] {$\l$} ++({\e-(\s)+1},0);
			}
			
			\foreach \s/\e/\l in {%
					2/\Rone-1/I_1,\Rone+3/\Rtwo-1/I_2,\Rtwo+3/\Rthree-1/I_3,\Rthree+2/\n/I_4,%
					1/1/t_1,\Rone+1/\Rone+2/t_2,\Rtwo+1/\Rtwo+2/t_3,\Rthree+1/\Rthree+1/t_4%
				} {
				\draw[decorate,decoration={brace,mirror,amplitude=2}] (\s-.9,-.25) 
					-- node[below=1pt,scale=.8] {$\l$} ++({\e-(\s)+.8},0);
			}
			
		\end{scope}

		\begin{scope}[shift={(0,-11)},pin distance=.7em]
			\node at (\n/2,5.65) {\textbf{Continuous-Values World}} ;
			
			\begin{scope}[shift={(0,3)}]
				\draw[|-|] (0,0) -- ++(\n,0);
				\node at (5,1) {sample} ;
				\node at (0,.75) {0}; \node at (\n,.75) {1};
				\foreach \x in {\Pone,\Ptwo,\Pthree} {
					\filldraw[fill=black!30,thick] (\n*\x,0) circle (1.1*\ps)  ;
				}
				\foreach \x in {0.06,0.25,0.4,0.61,0.7,.97} {
					\fill[white] (\n*\x,0) circle (\ps)  ;
					\begin{scope}[shift={(\n*\x,0)}]
						\clip (0,0) circle (\ps)  ;
						\foreach \i in {0,0.1625,...,1} {
							\draw[black,very thin] (.5,\i) -- ++(-1,-1);
						}
					\end{scope}
					\draw (\n*\x,0) circle (\ps)  ;
				}
			\end{scope}

			\draw[|-|] (0,0) -- ++(\n,0);
			\node at (0,.75) {0}; \node at (\n,.75) {1};

			\foreach \x in {0.06,0.25,0.4,0.61,0.7,.97} {
				\begin{pgfonlayer}{background}
				\fill[fill=black!5] (\n*\x,0) ++ (-0.9*\ps,0) rectangle ++(1.8*\ps,3);
				\end{pgfonlayer}
				\fill[white] (\n*\x,0) circle (\ps)  ;
				\begin{scope}[shift={(\n*\x,0)}]
					\clip (0,0) circle (\ps)  ;
					\foreach \i in {0,0.1625,...,1} {
						\draw[black,very thin] (.5,\i) -- ++(-1,-1);
					}
				\end{scope}
				\draw (\n*\x,0) circle (\ps)  ;
			}
			
			\foreach \x in {.035,.1,.22,.31,.33,.48,.5,.55,.655,.68,.722,.77,.85,.9,.91,.95} {
				\fill (\n*\x,0) circle (0.8*\ps) ;
			}
			\foreach \x/\i in {\Pone/1,\Ptwo/2,\Pthree/3} {
				\begin{pgfonlayer}{background}
				\fill[fill=black!5] (\n*\x,0) ++ (-0.9*\ps,0) rectangle ++(1.8*\ps,3);
				\end{pgfonlayer}
				\filldraw[fill=black!30,thick] (\n*\x,0) 
					node[pin={[scale=.7,fill=white,inner sep=1.5pt]270:$P_\i$}] {} circle (1.1*\ps);
			} 
			\begin{pgfonlayer}{background}
			\foreach \x in {0,\Pone,\Ptwo,\Pthree,1} {
				\draw[dotted] (\n*\x,-\ps) -- ++(0,-2.5);
			}
			\end{pgfonlayer}
			
			\node[scale=.7,anchor=east,inner sep=0pt] (p0) at (0,0.75) {$P_0={}~$} ;
			\node[scale=.7,anchor=west,inner sep=0pt] (p4) at (\n,0.75) {$~{}=P_4$} ;
			\filldraw[fill=black!20,opacity=.3] (0,0) circle (\ps)  (\n,0) circle (\ps);
			\draw[help lines,shorten <=3pt] (0,0) -- (p0.240);
			\draw[help lines,shorten <=3pt] (\n,0) -- (p4.-60);
			
			\foreach \f/\t/\i in {0/\Pone/1,\Pone/\Ptwo/2,\Ptwo/\Pthree/3,\Pthree/1/4} {
				\draw[<->,shorten >=1pt,shorten <=1pt] (\n*\f,-2) -- 
					node[scale=.8,fill=white,inner sep=1pt] {$D_\i$} ++({\n*(\t-\f)},0);
			}
			
		\end{scope}
	\end{tikzpicture}%
	}
	\caption{%
		Illustration of the notations used in the analysis;
		\textbf{top:} quantities that refer to counts 
		(sizes of segments and ranks of elements),
		\textbf{bottom:} quantities referring to numerical values of certain elements
		in the model of sorting $n$ \iid $\uniform(0,1)$ distributed numbers.%
	}
	\label{fig:notations}
	\vspace{-1.5ex}
\end{figure}
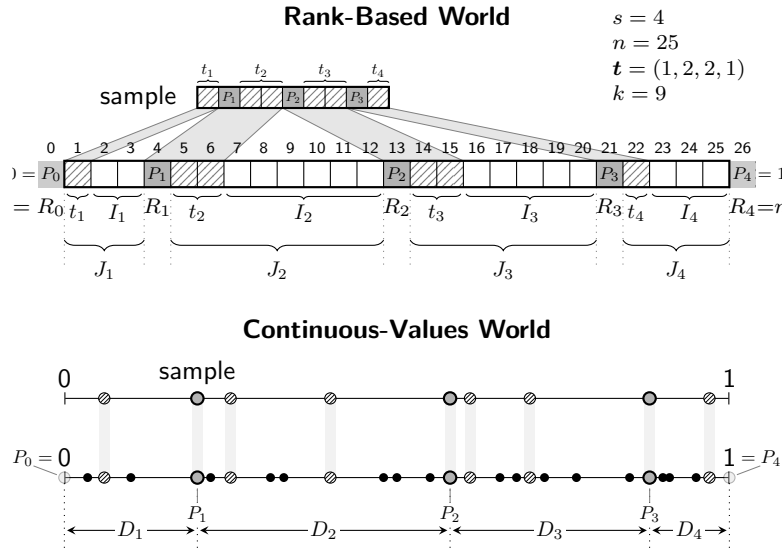

We can write the costs with $F_n \ce F_{n,M_n}$ as
\begin{align}
\label{eq:dist-rec-random-rank}
		F_n
	&\wwrel\eqdist
		A_{\mathcal F}(n) \bin+
		\sum_{\ell=1}^s \indicator{ \ui{R_{\ell-1}}n < M_n < \ui{R_{\ell}}n } \cdot \ui{F_{\ui{J_\ell}n}}{\ell}
		,
\end{align}
where $(\ui{F_j}\ell)_{j\in\N}$ are independent copies of $({F_n})_{n\in\N}$ for $\ell=1,\ldots,s$,
which are also independent of $\ui{\vect J}n$, 
$A_{\mathcal F}(n)$ and $M_n$.
This distributional recurrence is of the shape required for 
the distributional master theorem (DMT) 
to compute asymptotic approximations for the expected values.
We next check the technical conditions for the DMT.

\paragraph{Distribution of Subproblem Sizes}
\label{sec:distribution-subproblem-sizes}
For a single pivot chosen randomly
(without pivot sampling) we have $J_\ell \eqdist \uniform[0..n-1]$, a discrete uniform distribution.
In general, we have two summands: $J_\ell = t_\ell + I_\ell$. 
The first one accounts for the part of the sample that belongs to the $\ell$th subproblem,
which is a deterministic contribution dictated by the sampling scheme.
$I_\ell$ is the number of elements that were not part of the sample and
were found to belong to the $\ell$th segment in the partitioning step.
$I_\ell$ is a random variable, and its distribution is 
$I_\ell\eqdist \betaBinomial(n-k,t_\ell+1,k-t_\ell)$,
a so-called \textsl{beta-binomial distribution}.

The connection to the beta distribution is best seen by assuming $n$ independent and uniformly in $(0,1)$
distributed reals as input. They are almost surely pairwise distinct and their relative
ranking is a random permutation of~$[n]$, so this assumption is \withoutlossofgenerality
for our analysis.
Then, the $\ell$th subproblem contains all elements between $P_{\ell-1}$ and $P_\ell$,
$\ell=1,\ldots,s$.
The \emph{spacing} $D_\ell \ce P_\ell - P_{\ell-1}$ has a 
$\betadist(t_\ell+1,k-t_\ell)$ distribution (by definition!),
and \emph{conditional} on that \emph{spacing} $I_\ell \eqdist \binomial(n-k,D_\ell)$ 
has a binomial distribution:
Once the pivot values $P_{\ell-1}$ and $P_\ell$ are fixed,
any element falls between them with probability $D_\ell$.
The resulting mixture is the so-called beta-binomial distribution.
Note that for $\vect t=0$,
$t_\ell + \betaBinomial(n-k,t_\ell+1,k-t_\ell) = \betaBinomial(n-1,1,s-1)$
which coincides with $\uniform[0..n-1]$ for $s=2$.

\paragraph{Convergence of Relative Subproblem Sizes}
In light of the stochastic representation of the beta-binomial distribution,
we know that conditional on $D_\ell$, $Z_\ell = \ui{I_\ell}n / n $ is 
concentrated around~$D_\ell$.
Bounding the errors carefully yields
the required local limit law for the relative subproblem size $Z_\ell$:
By \wref{lem:local-limit-law-dirichlet-multinomial}, we find that
\weqref{eq:DMTwc-condition} is satisfied with $\delta=1$ and
the limiting density $f_{Z_\ell^*}(z) = z^{t_\ell}(1-z)^{k-t_\ell-1} / \BetaFun(t_\ell+1,k-t_\ell)$, 
which is the density of the $\betadist(t_\ell+1,k-t_\ell)$ distribution. 
$f_{Z_\ell^*}$ is clearly Lipschitz (and hence Hölder) continuous on $[0,1]$ 
since its derivative is bounded in $[0,1]$,
so the conditions of the DMT are fulfilled.


\paragraph{Conditional Convergence of Coefficients}
For the second condition, \weqref{eq:DMTwc-condition-coeffs}, we have to consider 
the distribution of $[R_{\ell-1} < M_n < R_{\ell}]$ conditional on 
the relative subproblem size $\ui{J_\ell}n / n$ for the $\ell$th recursive call.
Since $M_n$ is uniformly distributed, only the number of choices for $M_n$ in the considered range 
$[R_{\ell-1} < M_n < R_{\ell}]$
is important: $R_{\ell} - R_{\ell-1} - 1 = J_\ell$.
So we have
\(
		\Prob[\big]{ \ui{R_{\ell-1}}n < M_n < \ui{R_{\ell}}n \given \ui{J_\ell}n }
	\wwrel=
		\frac{\ui{J_\ell}n}n,
\)
so \weqref{eq:DMTwc-condition-coeffs} is fulfilled with $a_\ell(z) = z$ and $\delta=1$.
$f_{Z_\ell^*}(z) \cdot a_\ell(z)$ is Lipschitz and hence Hölder-continuous as required.


\paragraph{Solution for linear toll functions}
We can hence apply the master theorem.
The $Z_\ell^*$ from \wref{thm:DMTwc} correspond exactly to our spacings 
$D_\ell \eqdist \betadist(t_\ell+1,k-t_\ell)$.
We have
$\E{A_{\mathcal F}(n)} \sim  a_{\mathcal F}\cdot n$ with $a_{\mathcal F}$
a constant depending on the method and~$\mathcal F$.
So $\alpha=1$ and $\beta=0$ and
we compute 
\begin{align*}
		H
	&\wrel=
		1 - \sum_{\ell=1}^s \E[\big]{D_\ell^\alpha \,a_\ell(D_\ell)} 
	\wrel= 
		1 - \sum_{\ell=1}^s \E[\big]{D_\ell^2}
\\[-.5ex]	&\wrel= 
		1 - \sum_{\ell=1}^s \frac{ (t_\ell+1)^{\overline 2} } { (k+1)^{\overline 2} }
		.
\end{align*}
Since $\frac{ (t_\ell+1)^{\overline 2} } { (k+1)^{\overline 2} } < \frac{ t_\ell+1 } { k+1 }$
and $\sum_{\ell=1}^s  \frac{ t_\ell+1 } { k+1 } = 1$, we have $H>0$.
So by Case~1, we find
\(
		\E{F_n}
	\wwrel\sim
		\frac{a_{\mathcal F}}{H} \cdot n .
\)

%
%
%
%
%
%
%
%
%
%
%
%
%
%
%
%
%
%
%
%
%
%
%
%
%
%
%


\subsection{Generic Multiway Partitioning}
\label{sec:random-ranks-generic-sway}

The partitioning methods of practical relevance~--
classic Hoare-Sedgewick partitioning ($s=2$)~\cite{Sedgewick1978},
Lomuto partitioning~\cite{Bentley1984} ($s=2$),
YBB partitioning ($s=3$)~\cite{javacoredevel2009} and
``Waterloo partitioning'' ($s=4$)~\cite{Kushagra2014}~--
have been generalized to arbitrary $s$ and analyzed in \cite[Chapter\,5]{Wild2016}
with respect to the expected number of comparisons, scanned elements and write accesses.
By using the respective values for $a_{\mathcal F}$ 
given in \cite[\href{https://www.wild-inter.net/publications/html/wild-2016.pdf.html\#pff4}{Thm.\,7.1}]{Wild2016}
in the asymptotic expression for $\E{F_n}$, we obtain the overall costs for selecting random ranks;
(see \wref{tab:grand-average-no-sampling} for some results).

\subsection{Discussion}
\label{sec:random-ranks-discussion}

Although the results for generic $s$-way partitioning are readily available 
we refrain from stating them in full generality since 
the expressions are lengthy 
and many variants are not promising for selection.
Intuitively, a large $s$ can hardly be useful when we always recurse 
into only a single subproblem. 

\begin{table}[tbh]
	\smaller[1]
	\plaincenter{
	\begin{tabular}{>{\(}l<{\)} l *{3}{>{\(}l<{\)}}}
	\toprule
		s & Name     & \E{C_n}/n            & \E{\mathit{SE}_n}/n  & \E{\mathit{WA}_n}/n   \\
	\midrule
		2 & classic  & 3                    & 3                    & 1                     \\
		3 & YBB      & 3.1\overline6        & 2.\overline 6        & 1.8\overline 3        \\
		4 & Waterloo & 3.\overline3         & 2.5                  & 2                     \\
		5 &          & 3.5                  & 2.7                  & 2.35                  \\
		6 &          & 3.\overline6         & 2.8                  & 2.5\overline3         \\
		7 &          & 3.7\overline{857142} & 3.\overline{047619}  & 2.8\overline3         \\
		8 &          & 3.\overline{857142}  & 3.2\overline{142857} & 3.03\overline{571428} \\
	\bottomrule
	\end{tabular}
	}
	\caption{%
		The coefficient of the linear term of the expected number of comparisons,
		scanned elements and write accesses to the array
		for Quickselect with $s$-way partitioning without pivot sampling ($\vect t = 0$)
		when searching a random rank 
		(``grand averages'').
	}
	\label{tab:grand-average-no-sampling}
\end{table}

\oldparagraph{The optimal number of pivots}
\wref{tab:grand-average-no-sampling} confirms this intuition;
indeed for the classical cost measures of key comparisons ($C_n$) and
write access ($\mathit{WA}_n$, related to key exchanges)
there is no improvement whatsoever from multiway partitioning in Quickselect
(as pointed out before~\cite{WildNebelMahmoud2016}).
In terms of scanned elements ($\mathit{SE}_n$), however,
significant savings are observed.
Here, \wref{tab:grand-average-no-sampling} contains a surprise:
the minimum for scanned elements is attained for $s=4$!
This is against the intuition since 
there will always be (at least) two adjacent segments whose subdivision was fruitless.
How can this possibly be better than avoiding the extra work to produce a fourth segment?

The answer is that $s=4$ is indeed suboptimal; 
but our comparison in \wref{tab:grand-average-no-sampling} is not quite fair.
We do not select pivots from a sample, but the multiway methods do have to sort their $s-1$ 
pivots to operate correctly.
We therefore allow multiway methods to enjoy pivots of better quality
compared to methods with smaller $s$, thus giving the former an undue advantage.
This unfairness is inherent in any such comparison 
(as previously noticed in the context of sorting~\cite{Wild2016,Wild2018b}).

\paragraph{Simulation by binary partitioning}
A fair evaluation of the usefulness of multiway partitioning is nevertheless possible 
by considering the following (hypothetical) Quickselect variant:
We select pivots as we would for the $s$-way method, but then use 
\textsl{several rounds of classic single-pivot partitioning} to obtain the same
segments as with one round of the $s$-way method.
For example, the four segments produced by Waterloo partitioning could also
be obtained by first partitioning around the middle pivot and then 
the resulting left resp.\ right segment around the small resp.\ large pivot.
Note that the first round uses the median of three elements (the middle pivot),
whereas the second round effectively runs with pivots selected uniformly from their subrange.
By comparing the cost of both variants, we truly evaluate the quality of the partitioning methods 
since they use the same pivot values.

Comparing Waterloo partitioning with its simulation,
we observe that both execute exactly the same set of comparisons,
but \wrt scanned elements, the simulation scans each element twice. 
Waterloo partitioning scans all elements once and \textsl{only the elements in the outer two
segments a second time} (an average of $1.5n$ vs.\ $2n$ scanned elements).
This clearly exposes the superiority of multiway partitioning in terms of cache behavior
and explains its advantage for sorting.

In Quick\emph{select,} we will only pursue one subproblem recursively.
The simulation of Waterloo partitioning subdivides \emph{both} segments resulting from the 
first split, even though one will be knowingly useless!
We should therefore compare Waterloo select to a binary simulation 
without the useless second subdivision.
The number of scanned elements then is $n$
for the first round, plus the size of the segment on which we apply the second subdivision.
The probability to subdivide the left resp.\ right resulting segment 
is the relative size of that segment 
(the probability that the random rank lies there).
The partitioning costs are hence $n+\E{(J_1+J_2)^2/n} + \E{(J_3+J_4)^2/n} \sim 1.6n$ (for $\vect t=(0,0,0,0)$),
and the total cost are given by $\mathit{SE}_n \sim 1.6n / H(0,0,0,0) = 2.\overline6$.
This is still higher than $2.5n$, but much closer than $3n$.
That Waterloo-select performs so much better than classic Quickselect according to 
\wref{tab:grand-average-no-sampling}
is thus mostly due to the use of a median-of-3 pivot for the first partitioning round,
and only to a smaller extent due to its inherent advantage in terms of scanned elements.

We next consider YBB partitioning. 
Its simulation first partitions around the larger pivot and 
then subdivide the left segment around the smaller pivot.
This ``atomic'' version would incur $3.\overline 3n$ scanned elements, 
much more than the $2.\overline 6n$
of YBB-Select and indeed more than the $3n$ for classic Quickselect.
But for the lucky case that the sought pivot falls into the rightmost segment, 
the second subdivision is not needed and should be skipped;
this lazy version incurs on average $n+\E{(J_1+J_2)^2/n} \sim 1.5 n$ scanned elements per partitioning step
and thus still $1.5n / H(0,0,0) = 3n$ scanned elements in total.

\oldparagraph{Two pivots are optimal!}
But how does YBB-select compare to Waterloo-select?
A simulation of one by the other does not seem sensible, but
we can use the pivots for Waterloo-select (three random elements in order, $\vect t=(0,0,0,0)$)
in YBB-select. 
Ignoring the largest pivot and doing the three-way split using YBB-partitioning
corresponds to YBB-select with $\vect t=(0,0,1)$, which needs $\sim 2.5n$ scanned elements, 
the \emph{same} as Waterloo-select.

This statement is also true when we let Waterloo-select choose pivots equidistantly from a sample:
If $\vect t=(t,t,t,t)$ (for any $t\in\N_0$), the expected number of scanned elements
is $\sim \frac{4t+5}{2t+2} n$. Selecting pivots the same way, discarding the largest and using
YBB-partitioning with the two smaller pivots yields the same asymptotic result.
Of course, Waterloo-select performs more comparisons and array accesses to achieve this,
so we can conclude that when scanned elements dominate costs,
\textsl{dual-pivot partitioning is the unique optimum choice for Quickselect!}

\paragraph{Summary}
Splitting the input into several segments at the same time saves memory transfers.
While this unconditionally helps in sorting,
the game is different in selection where only one subproblem is considered recursively.
The flexibility to postpone the decision which part of the input should be further partitioned
(and hence the possibility to avoid the splitting of any discarded segments)
outperforms the savings in scanned elements from multiway partitioning.
Dual-pivot partitioning is an exception, though, since all splits were useful 
when we recurse into the middle segment.

\subsection{Adaptive Methods}
\label{sec:random-ranks-why-not-adaptive}

All the methods discussed above are non-adaptive: they only take the value of $m$
into account when they decide which subproblem to recurse into.
Unfortunately, this is an inherent limitation of the single-parameter recurrence that we use.
The validity of the recurrence relies on randomness preservation for $m$:
apart from the which subproblem contains the $m$th smallest element,
nothing has been learned about the rank of this element \emph{within} the subproblem.
Conditioned on the event that the sought rank is found in the given subproblem,
its rank is still uniformly distributed within the subproblem.

For adaptive methods, this is different.
Since partitioning costs and subproblem size distribution depend on the $v$ 
for which $\alpha \in I_v$, we inevitably learn which interval $\alpha$ lies in, 
in addition to the index of subproblem on which we recurse.
So even if $m$ is originally uniformly distributed in $[n]$, 
for recursive calls it is known to lie in a smaller range.
The grand average costs of adaptive Quickselect hence do not follow a simple 
one-parameter recurrence.

\section{Asymptotic Approximation for Linear Ranks}
\label{sec:linear-ranks-convergence}

We now consider selecting a fixed $\alpha$-quantile, where $\alpha\in(0,1)$
is a parameter of the analysis.
We start with the distributional equation \wref{eq:dist-rec-general}
and take expectations on both sides.
Since
we expect the overall costs to be asymptotically linear,
we divide by $n$:
\begin{multline*}
		\frac{\E{F_{n,m}}}n
	\wwrel=
		\frac{\E{A_{\mathcal F}(n,m)}}n
		\wbin+ \mkern-17mu\sum_{1\le \underline r < \overline r \le n} \mkern-10mu
			\frac{\overline r - \underline r - 1}{n} \times{}\\
			\Biggl( \sum_{\ell=1}^s \Prob[\big]{(R_{\ell-1},R_{\ell}) = (\underline r, \overline r) } \Biggr)
				\cdot \frac{\E{F_{\overline r - \underline r - 1,m - \underline r}}}{\overline r - \underline r - 1}
		.
\end{multline*}
\wref{thm:convergence} below confirms (under very general conditions) that
passing to the limit in this recurrence yields the desired asymptotic approximation.

This has been proven for single-pivot Quickselect 
even in a stochastic sense~\cite{Gruebel1999,GruebelRoesler1996};
the used techniques can be extended to adaptive methods
as outlined in~\cite{MartinezPanarioViola2010}.
\ifsiam{%
	We give an elementary proof that covers generic $s$-way Quickselect
	in the \extendedversion.
	Interestingly does not seem to appear in the literature.
	We point out that the computations are a bit lengthy,
	but do not need any sophisticated machinery:
}{%
	We give an elementary proof that covers generic $s$-way Quickselect
	and interestingly seems not to appear in the literature.
	The details are a bit lengthy and deferred to \wref{app:convergence}, 
	but we do not need any sophisticated machinery:
}
We simply use the \textsl{ansatz} $\E{F_{n,m}}\sim f(\frac mn)n$
to obtain an educated guess for $f$ and bound the error $\bigl|\E{F_{n,m}} - f(\frac mn)n\bigr|$. 
The latter fulfills a similar recurrence as $\E{F_{n,m}}$,
but with a much smaller toll function. 
A crude bound suffices for the claim.

\begin{theorem}[Convergence Linear Ranks]
\label{thm:convergence}
	Consider generic (adaptive) Quickselect 
	(as defined in \wref{sec:adaptive-quickselect}) and
	assume 
	$\E{A_\mathcal F(n,m)} = a_{\mathcal F}(\frac mn)n \bin\pm \Oh(1)$.
	Let $f:[0,1]\to\R_\ge 0$ be a function that fulfills the 
	following integral equation:
	\begin{align*}
		&	f(\alpha)
		\wrel=
			a_\mathcal F(\alpha) 
	\numberthis\label{eq:general-integral-equation}
	\\*[-.2ex]	&\ppe
				\bin+ \frac1{\BetaFun\bigl((t_1,\underrightarrow{t_1})+1\bigr)}
					\int_{u=\alpha}^1 \mkern-10mu u^{t_1+1} (1-u)^{\underrightarrow{t_1}} 
						f\Bigl(\frac\alpha u\Bigr) \, du
	\\	&\ppe
				\bin+ \frac1{\BetaFun\bigl((t_s,\underleftarrow{t_s})+1\bigr)}
					\int_{v=0}^\alpha \mkern-10mu v^{\underleftarrow{t_s}} (1-v)^{t_s+1} 
						f\Bigl(\frac{\alpha-v} {1-v}\Bigr) \, dv
	\\	&\ppe
				\bin+ \sum_{\ell=2}^{s-1}
					\frac1{\BetaFun\bigl((\underleftarrow{t_\ell},t_\ell,\underrightarrow{t_\ell})+1\bigr)}
					\times{}
	\\*	&\ppe
					\int_{u=0}^\alpha\int_{v=\alpha}^1 \mkern-10mu
						u^{\underleftarrow{t_\ell}} (v-u)^{t_\ell+1} (1-v)^{\underrightarrow{t_\ell}} 
						f\Bigl(\frac{\alpha-u}{v-u}\Bigr) \, dv\,du,
	\end{align*}
	where we abbreviate
	\(
	 		\underleftarrow{t_\ell}
		\wwrel=
			\sum_{r=1}^{\ell-1} (t_r + 1) \bin-1
	\text{ and } 
			\underrightarrow{t_\ell}
		\wwrel=
			\sum_{r=\ell+1}^{s} (t_r + 1) \bin-1
	\).
	For adaptive methods, $s$ and $\vect t$ are functions of $\alpha$, 
	which is suppressed for legibility.
	Then \wref{eq:general-integral-equation} is required piecewise for $\alpha\in I_v$, $v\in[d]$.
	
	Assume that $f$ is ``(piecewise) smooth'', \ie, $f$ (restricted to $I_v$) is 
	Hölder-continuous with exponent $h\in(0,1]$ (for all $v\in[d]$).
	Then the limit
		\(\lim_{n\to\infty; \frac mn \to\alpha} {\E{F_{n,m}}}/{n}\)
	exists for $\alpha\in(0,1)\setminus\mathcal A$,
	where $\mathcal A$ is the set of boundaries 
	of $I_1,\ldots,I_d$.
	
	Moreover, with $m=\lceil\alpha n\rceil$ holds
	\begin{align*}
			\E{F_{n,m}}
		&\wwrel=
			f(\tfrac mn) n \wbin\pm \Oh(n^{1-2h/3})
			,\quad (n\to\infty).
	\end{align*}
\end{theorem}

Our continuity requirements for $f$ may appear restrictive, 
but they are fulfilled in all examples we studied.
They might indeed follow from \wref{eq:general-integral-equation} in general,
but we do not attempt to prove this conjecture.

\paragraph{\boldmath How to obtain $f$}
\weqref{eq:general-integral-equation} determines $f$ only implicitly.
Our route to an explicit expression consists of the following steps.%
\begin{inlineenumerate}
\item 
	Use substitutions to obtain integrals that only involve $f(x)$ 
	instead of the shifted and scaled arguments.
\item
	Take successive derivatives on both sides until all integrals vanish.
	This will result in a higher-order differential equation for $f$ that we aim to solve.
\item
	Compute $f$ by determining constants of integration from boundary conditions
	and known results (\eg, symmetry and results for $\alpha\to0$ and random ranks).
\end{inlineenumerate}

\paragraph{Separable equations for adaptive sampling}
	Since taking derivatives does not change the argument of $f$, 
	the differential equation will only relate different derivatives of $f$ 
	evaluated \textsl{at the same point~$x$}.
	This is vital for adaptive sampling since it means that we can solve the differential
	equation for each $I_v$ separately.
	Only step 3) involves the interactions of the regimes.

We remark that we can obtain the leading-term coefficient of
the grand average by integrating:
\(
		\bar f 
	\wwrel=
		\int_0^1 f(\alpha) \,d\alpha
\).
This also works for adaptive methods.

The discussion in \wref{sec:random-ranks} justifies a restriction to $s \le 3$ segments,
but an explicit solution for the differential equation seems out of reach
for the general case.
We therefore focus on the simplest special cases first.

\section{YBB-Select with Linear Ranks}
\label{sec:linear-ranks-ybb}

As a warm-up, and part of our main result on Sesquickselect,
we consider YBB-Select (YQS) without sampling (as studied in~\cite{WildNebelMahmoud2016}).
We start with \wref{eq:general-integral-equation} and substitute 
$x\mapsto \alpha/u$, $x\mapsto \frac{\alpha-v}{1-v}$ and $x\mapsto \frac{\alpha-u}{v-u}$
in the first, second and third integral, respectively.
Simplifying the integrals is fairly standard; 
we show details for the most interesting one:
\begin{align*}
	&\mkern-0mu
		\int_{v=0}^\alpha \int_{v=\alpha}^1 
		(v-u) f\Bigl(\frac{\alpha-u}{v-u}\Bigr) \, dv\, du
\\	&\wrel=
		\int_{v=0}^\alpha (v-\alpha)^2 \int_{x=0}^{\alpha/v} \frac{f(x)}{(1-x)^3} \,dx\,dv
\\	&\wrel=
		\int_{x=0}^1 \frac{f(x)}{(1-x)^3} \int_{v=\alpha}^{\min\{1,\frac\alpha x\}} (v-\alpha)^2 \,dv\,dx
\\	&\wrel=
		\frac{(1-\alpha)^3}{3} \mkern-4mu\int_{x=0}^\alpha \frac{f(x)}{(1-x)^3} \,dx
		\bin+
		\frac{\alpha^3}{3} \mkern-3mu\int_{x=\alpha}^1 \mkern-7mu\frac{f(x)}{x^3} \,dx
		.
\end{align*}
Inserting yields the integral equation for $\vect t=(0,0,0)$,
\begin{align*}
		f(\alpha) 
	&\rel=
		a_\mathcal{F}(\alpha) 
		\bin+ 
\numberthis\label{eq:integral-eq-yqs}
\\*	&\rel\ppe
		2\Biggl(
			\alpha^2\int_\alpha^1\frac{f(x)}{x^3}\,dx
			\bin-\alpha^3\int_\alpha^1\frac{f(x)}{x^4}\,dx
\\*	&\rel\ppe
			\bin+ (1-\alpha)^2 \mkern-4mu \int_0^\alpha \mkern-7mu \frac{f(x)\,dx}{(1-x)^3}
			- (1-\alpha)^3 \mkern-4mu \int_0^\alpha \mkern-7mu \frac{f(x)\, dx}{(1-x)^4}
\\*	&\rel\ppe
			\bin+ \frac{(1-\alpha)^3}{3} \mkern-5mu \int_0^\alpha \mkern-7mu \frac{f(x)}{(1-x)^3}\,dx
			\bin+ \frac{\alpha^3}{3} \mkern-3mu \int_\alpha^1 \mkern-3mu \frac{f(x)}{x^3}\,dx
		\Biggr),
\end{align*}
and taking derivatives four times yields
\begin{equation}
		\frac{d^4f}{d\alpha^4}
	\wwrel=
		\frac{d^4 a_\mathcal{F}}{d\alpha^4}
		\bin+ 2\cdot \frac{1-3\alpha(1-\alpha)}{\alpha^2(1-\alpha)^2}\cdot\frac{d^2f}{d\alpha^2}
		.
\label{eq:deq-dual-pivot-no-sampling}
\end{equation}
For comparisons $a_\mathcal{C}(\alpha)=19/12$ and for scanned elements $a_\mathcal{SE}(\alpha)=4/3$. 
More generally, if 
$a_\mathcal{F}(\alpha)=a$ for some constant $a$ we can solve~\eqref{eq:deq-dual-pivot-no-sampling} to get
\begin{align*}
		f(\alpha) 
	&\wrel= 
		C_1
		+C_2\cdot\alpha
\\*	&\wrel\ppe{}
		+C_3\cdot\left(1-(1-\alpha)\ln(1-\alpha)-\alpha\ln(\alpha)\right)
\\*	&\wrel\ppe{}
		+C_4\Bigl(
			\tfrac{3}{10}\alpha^5-\tfrac{3}{4}\alpha^4+\tfrac{1}{6}\alpha^3+\tfrac{1}{2}\alpha^2
\\*[-1ex]	&\wrel\ppe\quad\qquad{}
			-(1-\alpha)\ln(1-\alpha)+1-\alpha
		\Bigr)
\end{align*}
for some constants $C_i$, $i=1,\ldots,4$, to be determined. 
If $a_\mathcal{F}(\alpha)$ is symmetric, that is, 
$a_\mathcal{F}(\alpha)=a_\mathcal{F}(1-\alpha)$ 
for any $\alpha\in[0,1]$ then $f(\alpha)$ is also symmetric, and this
entails $C_2=C_4=0$. Therefore
\(
	f(\alpha) = C_1+C_3\cdot(1+h(\alpha))
\)
where $h(\alpha)=-(1-\alpha)\ln(1-\alpha)-\alpha\ln(\alpha)$.
We have $a_\mathcal{F}(\alpha)=a$ for some constant $a$,
and from \wref{sec:random-ranks}, we then know $\overline{f}=2a$.
Moreover, we can also determine $\E{F_{n,1}}$
with the DMT (see also~\cite{WildNebelMahmoud2016}) and
find
$f(0)=3a_\mathcal{F}(0)/2$ 
(this equality holds in terms of right limits when $\alpha\to 0^+$).
These two equations determine $C_1$ and $C_3$
and we obtain
\begin{equation}
f^{[\mathrm{YQS}]}(\alpha) \wrel= a^{[\mathrm{YQS}]} \left(\frac{3}{2}+h(\alpha)\right).
\end{equation}
Recall that for standard quickselect 
$f^{[\mathrm{CQS}]}(\alpha)=a^{[\mathrm{CQS}]}(2+2h(x))$ (\wref{sec:previous-work}).
We stress here that the values for $a$ are different for CQS and YQS.

\begin{figure}[tbh]
	\hspace*{-.5em}
		\includegraphics[width=.49\linewidth]{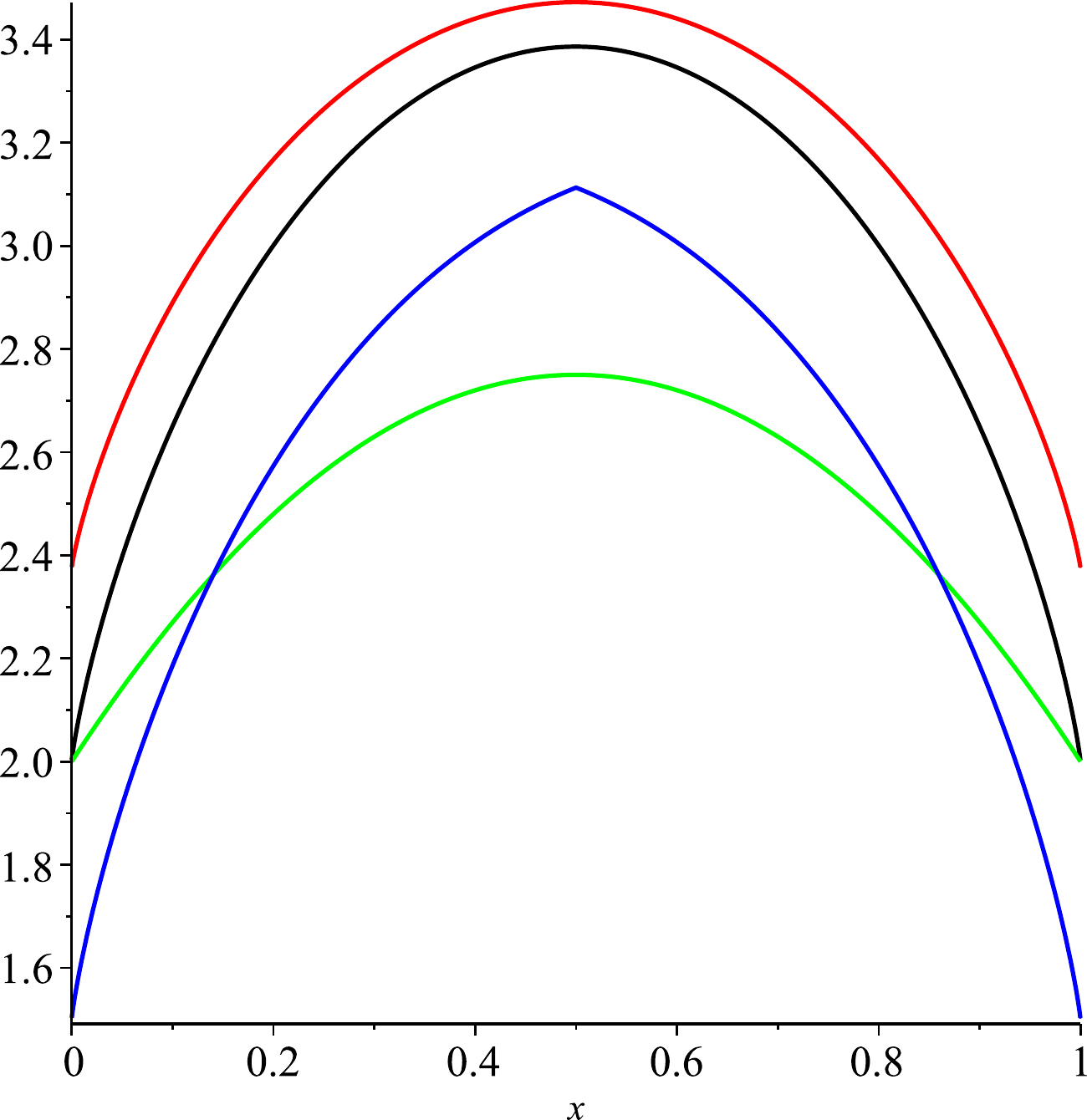}\hfill%
		\includegraphics[width=.49\linewidth]{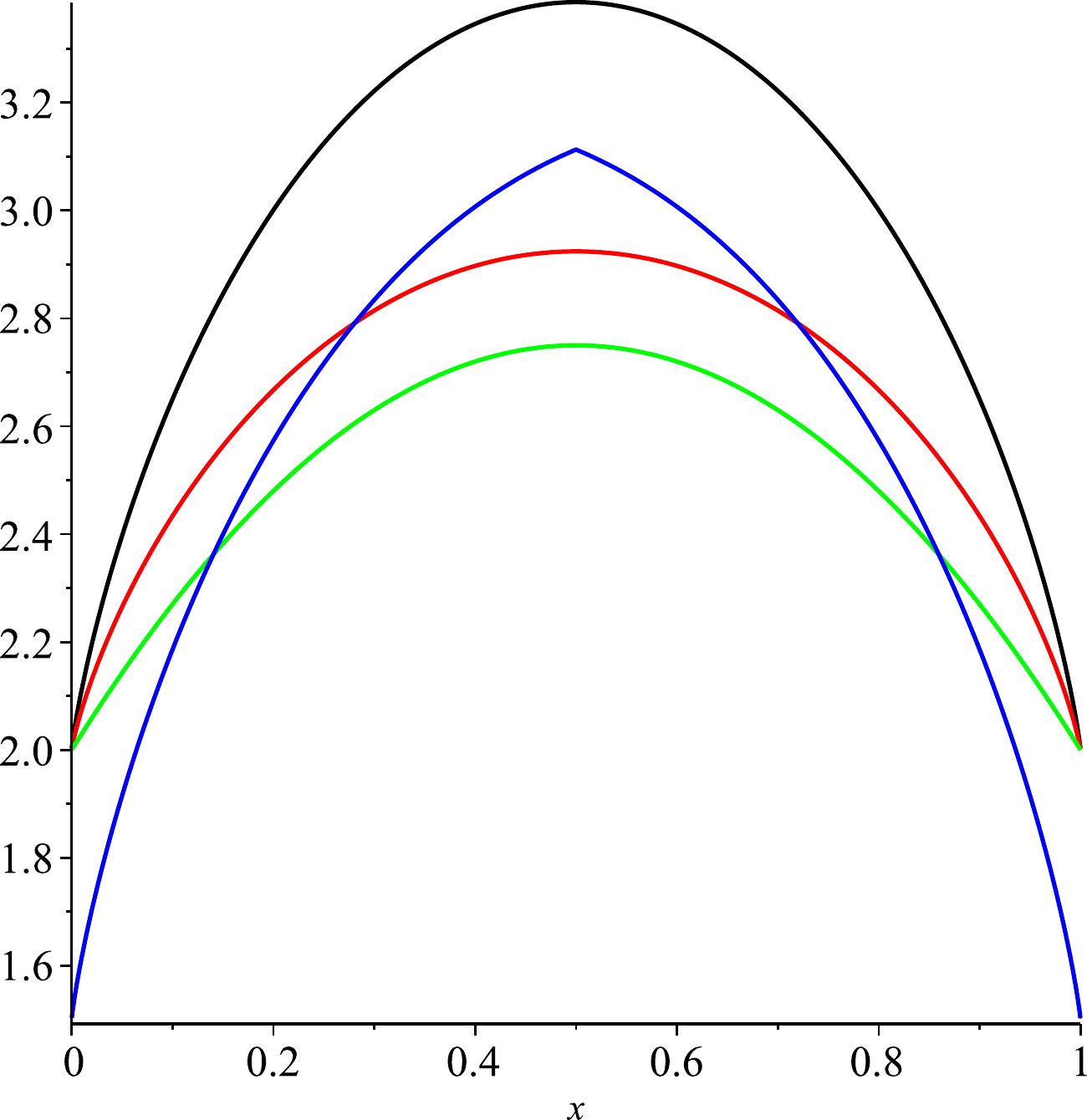}%
	\caption{%
		Key comparions, $c(\alpha)$, (\textbf{left})
		and 
		scanned elements, $\mathit{se}(\alpha)$, (\textbf{right})
		for standard Quickselect (black),
		YBB-select (red), median-of-three Quickselect (green) and proportion-from-2 
		(blue).
	}
	\label{fig:comparison-scanned-elements}
\end{figure}

\paragraph{Discussion}
Classic Quickselect (CQS) uses 
fewer comparisons than YBB-Select (YQS)
not only in the grand average, but
for \emph{any} fixed relative rank,
(see \wref{fig:comparison-scanned-elements} left). 
Similarly for write accesses, which we omit due to space constraints.
For scanned elements, however,
YQS scans \emph{less} elements on average 
than CQS for \emph{any} relative rank $\alpha$,
(\wref{fig:comparison-scanned-elements} right)! 
The difference
\(
\mathit{se}^{[\mathrm{CQS}]}(\alpha)-\mathit{se}^{[\mathrm{YQS}]}(\alpha)= \frac23 h(\alpha)
\)
is positive for all $\alpha\in(0,1)$ and reaches a maximum of about $13.6\%$ more
at $\alpha=1/2$ 
(approx.\ $3.386$ vs.\ $2.924$). 
As we know from \wref{sec:random-ranks} (\wref{tab:grand-average-no-sampling}), 
$\overline{se}^{[\mathrm{CQS}]}=3$ and $\overline{se}^{[\mathrm{YQS}]}=2.\overline6$, \ie, 
on average for random ranks, YQS scans $11.1\%$ less elements than CQS. 

\wref{fig:comparison-scanned-elements} shows two further
Quickselect variants.
Median-of-three Quickselect (M3) beats YQS on all ranks, 
but the comparison is not quite fair because of the larger sample.
\emph{Proportion-from-2} (PROP2), however, 
uses the same sample size as YQS:
It selects the smaller resp.\ larger of two sampled elements depending on whether
$\alpha\le 1/2$ or $\alpha > 1/2$ holds.%
\footnote{%
	\ifsiam{\strut}{}%
	The idea generalizes to proportion-from-$k$ (PROP$k$) for any sample size~$k$,
	where further cutoffs are introduced.
	In \emph{proportion-from-3}, for example,
	if $\alpha<\nu$
	for a parameter $\nu\in[0,\frac12]$,
	the smallest element of three elements is used as the pivot; 
	if $\alpha>1-\nu$, the largest element is used,
	and the median of the sample is used whenever 
	$\alpha \in [\nu,1-\nu]$. 
}
This adaptive variant was considered in~\cite{MartinezPanarioViola2010};
it is optimal \wrt comparisons for sample size~2
and beats YQS \wrt scanned elements for extremal~$\alpha$
(roughly when $\alpha\le 0.281$ or $\alpha\ge 0.719$)
and grand average
($\overline{\mathit{se}}^{[\mathrm{PROP2}]} \approx 2.598$).
The dual-pivot equivalent of proportion-from-2 that we study next
will improve on this significantly.

\section{Sesquickselect}   
\label{sec:sesquickselect}

Like PROP2, Sesquickselect (SQS) uses two sample elements.
If $\alpha<\nu$ for a parameter $\nu\in[0,\frac12]$, 
we use the smaller element in the sample to partition the array, 
if $\alpha > 1-\nu$ use the larger element, 
and if $\alpha \in [\nu,1-\nu]$ use \emph{both}
elements as pivots in YBB partitioning.
We now analyze Sesquickselect on linear ranks following
the same steps as in \wref{sec:linear-ranks-ybb}.

Provided $f(\alpha)=\lim_{n\to\infty,m/n\to\alpha}\Exp{F_{n,m}}/n$ 
exists (recall \wref{sec:linear-ranks-convergence}),
it will be a piecewise-defined function: 
$f(\alpha) = f_1(\alpha)$ for $\alpha\in I_1 \ce [0,\nu)$,
$f(\alpha) = f_2(\alpha)$ for $\alpha\in I_2 \ce [\nu,1-\nu]$, and
$f(\alpha) = f_3(\alpha)$ for $\alpha\in I_3 \ce (1-\nu,1]$.
Moreover, since $a_\mathcal{F}(\alpha)$ is symmetric 
(\ie, $a_\mathcal{F}(\alpha)=a_\mathcal{F}(1-\alpha)$) 
so is $f(\alpha)$, 
which implies $f_3(\alpha)=f_1(1-\alpha)$ and
$f_2(\alpha)=f_2(1-\alpha)$. 
Only two ``pieces'', say $f_1$ and $f_2$, thus have to be determined.
For $f_2(\alpha)$, we find that 
it satisfies the very same differential equation, \wref{eq:deq-dual-pivot-no-sampling}, as $f^{[\mathrm{YQS}]}$.
This is not surprising; $f_2$ is the YQS branch of SQS and
the differential equation only uses local properties of $f$ 
(as pointed out in \wref{sec:linear-ranks-convergence}). And inside $I_2$, $f=f_2$.

Likewise, $f_1(\alpha)$ satisfies the same differential equation 
as the function $f_1$ 
in PROP2 (see~\cite{MartinezPanarioViola2010}):
\begin{align*}
		\frac{d^4f_1}{d\alpha^4}
	&\wrel=
		\frac{d^4 a_\mathcal{F}}{d\alpha^4}
		\bin+\frac{2}{\alpha^2}
			\cdot \frac{d^2f_1}{d\alpha^2}
		\bin+\frac{2}{(1-\alpha)}
			\cdot \frac{d^3f_1}{d\alpha^3}
		.
\end{align*}
The derivatives of $a_\mathcal{F}(\alpha)$ vanish (inside any $I_v$), and 
we can reduce the order of both differential equations using 
$\phi_1=f_1''$ and $\phi_2=f_2''$.
This yields
\begin{align*}
		f_1(x)
	&\wrel= 
		C_1 \Bigl(
			\tfrac16 x^3
			+\tfrac12 x^2-x
			-(1-x)\ln(1-x)
		\Bigr)
\\*	&\wrel\ppe
		\bin+C_2h(x)
		\bin+C_3x 
		\bin+C_6,
\\
		f_2(\alpha)
	&\wrel= 
		C_4 \bin+ C_5h(x),
\end{align*}
for constants $C_1,\ldots,C_6$ that depend on the 
cost measure and threshold $\nu$.
We will write $f(\alpha) = f_\nu(\alpha)$ 
resp.\ $\nu$-SQS to stress this latter dependence.
The symmetry of $f_2$ was already taken into account.
Since $f_1(0)=\frac32 a_\mathcal{F}(0)$
we can eliminate $C_6 = \frac32 a_\mathcal{F}(0)$.
To determine the remaining constants, we insert the general expression for $f_1$ 
and $f_2$ into the integral equation and equate.
The process is laborious, but doable with computer algebra;
we report explicit expressions for $C_1,\ldots,C_5$ 
for comparisons and scanned elements in 
\ifsiam{%
	the \extendedversion.%
}{%
	\wref{app:constants-sqs}.%
}

\paragraph{Discussion}
We will focus on scanned elements.
To understand how $\nu$-SQS behaves for different~$\nu$,
we consider $g_1(\nu)=\lim_{\alpha\to\nu^-}f_{1,\nu}(\alpha)$ and 
$g_2(\nu)=\lim_{\alpha\to\nu^+}f_{2,\nu}(\alpha)$,
the values of the two branches of $f_\nu$ 
at $\nu$. 
We have 
$g_1(0)=\lim_{\nu\to 0^+}g_1(\nu)=1.\overline6$
and $g_2(0)=\lim_{\nu\to 0^+}g_2(\nu)=2$,
but
$g_1(\frac12)\approx 3.11$ and
$g_2(\frac12)\approx 2.91$
Since $g_1$ and $g_2$ are continuous 
and strictly increasing for $\nu \in (0,\frac12)$,
they cross at a unique point $\nu = \nu^\ast \in (0,\frac12)$.
This point is indeed the right choice:
\begin{theorem}[Optimal Sesquickselect]
\label{thm:sqs-optimal-nu}
	There exists an optimal value of $\nu^\ast\approx 0.265\,717$ such that 
	$se_{1,\nu^\ast}(\nu^\ast) = se_{2,\nu^\ast}(\nu^\ast)$.
	$\nu^*$-SQS
	scans fewer elements than other $\nu$-SQS, \ie,
	\(
	se_{\nu^\ast}(\alpha)\le se_\nu(\alpha), 
	\)
	for all $\nu\in [0,\frac12]$ and all $\alpha\in[0,1]$;
	in particular, 
	$se_{\nu^\ast}(\alpha)\le se^{[\mathrm{YQS}]}(\alpha)$ and 
	$se_{\nu^\ast}(\alpha)\le se^{[\mathrm{PROP2}]}(\alpha)$ 
	for any $\alpha\in[0,1]$.
\end{theorem}
The proof is similar to \cite[Thm~5.1]{MartinezPanarioViola2010},
we give the details 
\ifsiam{%
	in the \extendedversion.
}{%
	in \wref{app:opt-sqs-nu}.%
}%

Since $\nu^\ast$ is optimum across all relative ranks, 
$\nu^\ast$ minimizes $se_\nu(1/2)$ 
and $\overline{se}_\nu$: we have $se_{\nu^\ast}(1/2) \approx 2.843$ 
and $\overline{se}_{\nu^\ast} \approx 2.5004$,
(see also \wref{tab:special-values}).

\begin{table}[tbh]
	\smaller[1]
	\plaincenter{
		\begin{tabular}{rlllll}
		\toprule
			         &           & \multicolumn{2}{c}{Comparisons} & \multicolumn{2}{c}{Scanned elements} \\
			$\alpha$ & PROP2     & YQS             & $\nu^*$-SQS    & YQS            & $\nu^*-$SQS          \\
		\midrule
			     $0$ & $1.5$     & $2.375$         & $1.5$         & $2$            & $1.5$               \\
			   $1/2$ & $3.113^+$ & $3.472^+$       & $3.252^+$     & $2.924^+$      & $2.843^+$           \\
			     avg & $2.598^+$ & $3.1\overline6$ & $2.733^+$     & $2.\overline6$ & $2.500^+$           \\
		\bottomrule
			         &           &                 &               &                &
		\end{tabular}
	}%
	\vspace{-2ex}
\caption{%
	Some special values of $c(\alpha)$ ($\overline{c}$) and $se(\alpha)$ ($\overline{se}$) for several
	variants: YQS ($\nu=0$), PROP2 ($\nu=1/2$) and $\nu^*$-SQS.
	(Recall that $c^{[\mathrm{PROP2}]} = \mathit{se}^{[\mathrm{PROP2}]}$.)
}
\label{tab:special-values}
\end{table}

\subsection{Sesquickselect with larger samples}
\label{sec:sesquickselect-sampling}

The idea of Sesquickselect naturally extends to more than two sample elements:
SQS$k$ adaptively chooses one or two elements as pivot(s) from a sample of~$k$.
(SQS is simply SQS2 in this notation).
For larger~$k$, there are many options to do this and guidance is needed
to select good variants.
With pivot sampling, scanned-element costs for YBB partitioning are
$a_\mathcal{SE} = 1 + (t_1+1)/(k+1)$;
when $t_3 < t_1$, we can improve this to $1 + (t_3+1)/(k+1)$ using ``BBY partitioning'',
a symmetric variant of YBB partitioning.
We hence assume here that 
$a_\mathcal{SE} = 1 + \frac{\min\{t_1,t_3\}+1}{k+1}$.%

We could give a complete analysis for SQS2,
but for larger $k$ the higher-order differential equations 
seem to withstand analytic solutions.
We can, however, numerically solve the integral equation~\wref{eq:general-integral-equation}
to get insight into which adaptive variants are promising algorithms.
The code is available online: \url{https://github.com/sebawild/quickselect-integral-equation}.
Although numeric convergence was very good in all our explorations,
we do not prove the validity of the numeric procedures.
The smoothness requirement for \wref{thm:convergence}
seemed likewise to be fulfilled in all cases, 
but it remains a working hypothesis for this section.

We conjecture that the variants given in \wref{fig:sesquickselects} 
are the (approximately) optimal choices for the given sample size;
they have been found by extensive albeit non-exhaustive search.
\wref{fig:sesquickselects-se-alpha} compares the scanned-elements cost for Sesquickselect and 
biased proportion-from-$k$ for small~$k$.

\paragraph{Discussion}

We observe that for $k\le7$
we are still far away from the optimal leading term of $1+\min\{\alpha,1-\alpha\}$.
For example, $\overline{\mathit{se}}^{[\mathrm{SQS7}]} \approx 1.841$,
almost 50\% more than the optimal $1.25$.
This is quite different in sorting, where median-of-7 Quicksort 
is less than 10\% above optimum in the leading term.

A possible explanation is that 
the variance of the pivot ranks is too big.
Consider, \eg, $k=7$ and $\alpha\in[0.465,0.5]$.
The probability to recurse on the middle segment 
for, \eg, $\vect t=(2,0,3)$ is only roughly $1$\%.
We must therefore use rather balanced sampling vectors 
(here $\vect t=(1,2,2)$) and thus lose the ability
to reduce the problem size to much less than $\frac13 n$ in one step.

\tikzset{
	sep/.style={
		font={\smaller[2]},
		below,
	},
	tvec/.style={
		font=\smaller[2],
		above,
	},
}
\newboolean{showzeroone}
\setboolean{showzeroone}{true}
\newcommand\drawsesquickselect[2]{%
	\draw[thick] let \p{right} = (1,0) in 
		(0,0) node[xshift=-1em,yshift=-.25ex,anchor=east,] {#1}
		 -- ++(0,.1) -- ++(0,-.2) 
			node[sep] {\ifthenelse{\boolean{showzeroone}}{$0$}{}} 
		++(0,.1)
		foreach \da/\t in {#2} {
			-- node[tvec]{\t} ++(\da,0) 
			-- ++(0,.1) -- ++(0,-.2)  node[sep] {%
					\pgfmathparse{\currentx/\x{right}}%
					\ifthenelse{\boolean{showzeroone}}{%
						\pgfmathprintnumber{\pgfmathresult}%
					}{%
						\ifthenelse{\lengthtest{\currentx < \x{right}}}{%
							\pgfmathprintnumber{\pgfmathresult}%
						}{}%
					}%
				} ++(0,.1)
		}
	;
}

\begin{figure}[tbh]
	\ifarxiv{%
	}{
		\hspace*{-1em}
	}%
	\adjustbox{max width=.99\linewidth,center}{%
	\begin{tikzpicture}[
		xscale=11,
	]
		\def\yskip{1.35}
		\pgfkeys{/pgf/number format/.cd,fixed,precision=4,fixed zerofill=false}
		\begin{scope}[shift={(0,-1*\yskip)}]
		\drawsesquickselect{$k=3$}{
			0.1035/{(0,2)},
			{0.5-0.1035}/{(0,0,1)},
			{0.5-0.1035}/{(1,0,0)},
			0.1035/{(2,0)}
		}
		\end{scope}
		\begin{scope}[shift={(0,-2*\yskip)}]
		\drawsesquickselect{$k=4$}{
			0.06/{(0,3)},
			{0.28-0.06}/{(0,0,2)},
			{0.5-0.28}/{(0,1,1)},
			{0.5-0.28}/{(1,1,0)},
			{0.28-0.06}/{(2,0,0)},
			0.06/{(3,0)}
		}
		\end{scope}
		\tikzset{
			tvec/.style={
				font=\smaller[2],below,
				scale=.7,
				inner sep=1pt,
				anchor=south west,
				rotate=50,
			},
			sep/.style={
				font=\smaller[2],below,
				scale=.7,
				inner sep=1pt,
				anchor=west,
				rotate=-90,
			},
		}
		\begin{scope}[shift={(0,-3*\yskip)}]
		\drawsesquickselect{$k=5$}{
			0.036/{(0,4)},
			{0.153-0.036}/{(0,0,3)},
			{0.5-0.153}/{(0,1,2)},
			{0.5-0.153}/{(2,1,0)},
			{0.153-0.036}/{(3,0,0)},
			0.036/{(4,0)}
		}
		\end{scope}
		\begin{scope}[shift={(0,-4*\yskip)}]
		\drawsesquickselect{$k=6$}{
			0.025/{(0,5)},
			{0.09-0.025}/{(0,0,4)},
			{0.38-0.09}/{(0,1,3)},
			{0.5-0.38}/{(1,1,2)},
			{0.5-0.38}/{(2,1,1)},
			{0.38-0.09}/{(0,1,3)},
			{0.09-0.025}/{(4,0,0)},
			0.025/{(5,0)}
		}
		\end{scope}
		\begin{scope}[shift={(0,-5*\yskip)}]
		\drawsesquickselect{$k=7$}{
			0.02/{(0,6)},
			{0.06-0.02}/{(0,0,5)},
			{0.2875-0.06}/{(0,1,4)},
			{0.465-0.2875}/{(1,1,3)},
			{0.5-0.465}/{(1,2,2)},
			{0.5-0.465}/{(2,2,1)},
			{0.2875-0.06}/{(4,1,0)},
			{0.465-0.2875}/{(3,1,1)},
			{0.06-0.02}/{(5,0,0)},
			0.02/{(6,0)}
		}
		\end{scope}
		
	\end{tikzpicture}
	}
	\vspace{-1ex}
	\caption{%
		Our conjectured (approx.) optimal SQS$k$ variants for small $k$.
		If $\alpha$ falls in the interval (delimited by the values given below the line), 
		the vector above the line is used for $\vect t$.
		When $\vect t$ has two entries, classic partitioning is used;
		where three entries are given, we use YBB-partitioning for $\alpha \le \frac12$
		and BBY-partitioning for $\alpha>\frac12$.
	}
	\label{fig:sesquickselects}
\end{figure}
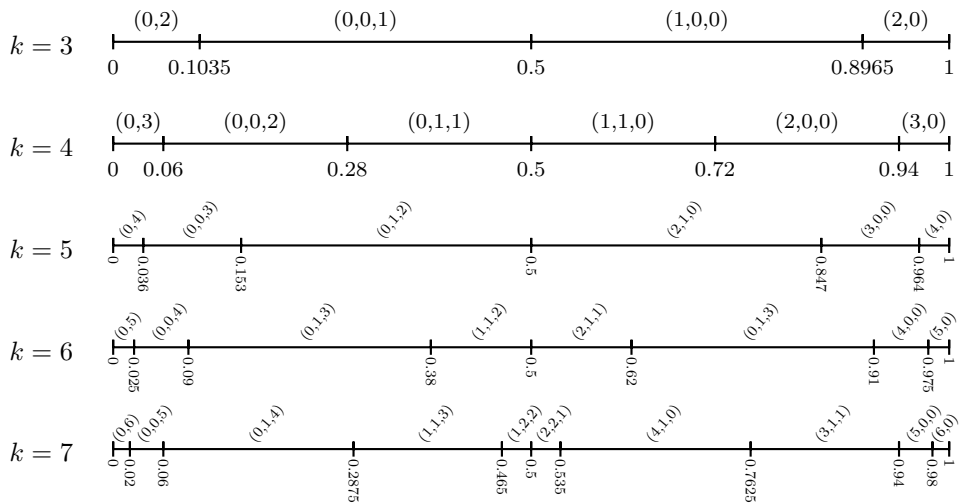

\begin{figure}[tbh]
	\plaincenter{\hspace*{-.5em}
		\includegraphics[width=.98\linewidth]{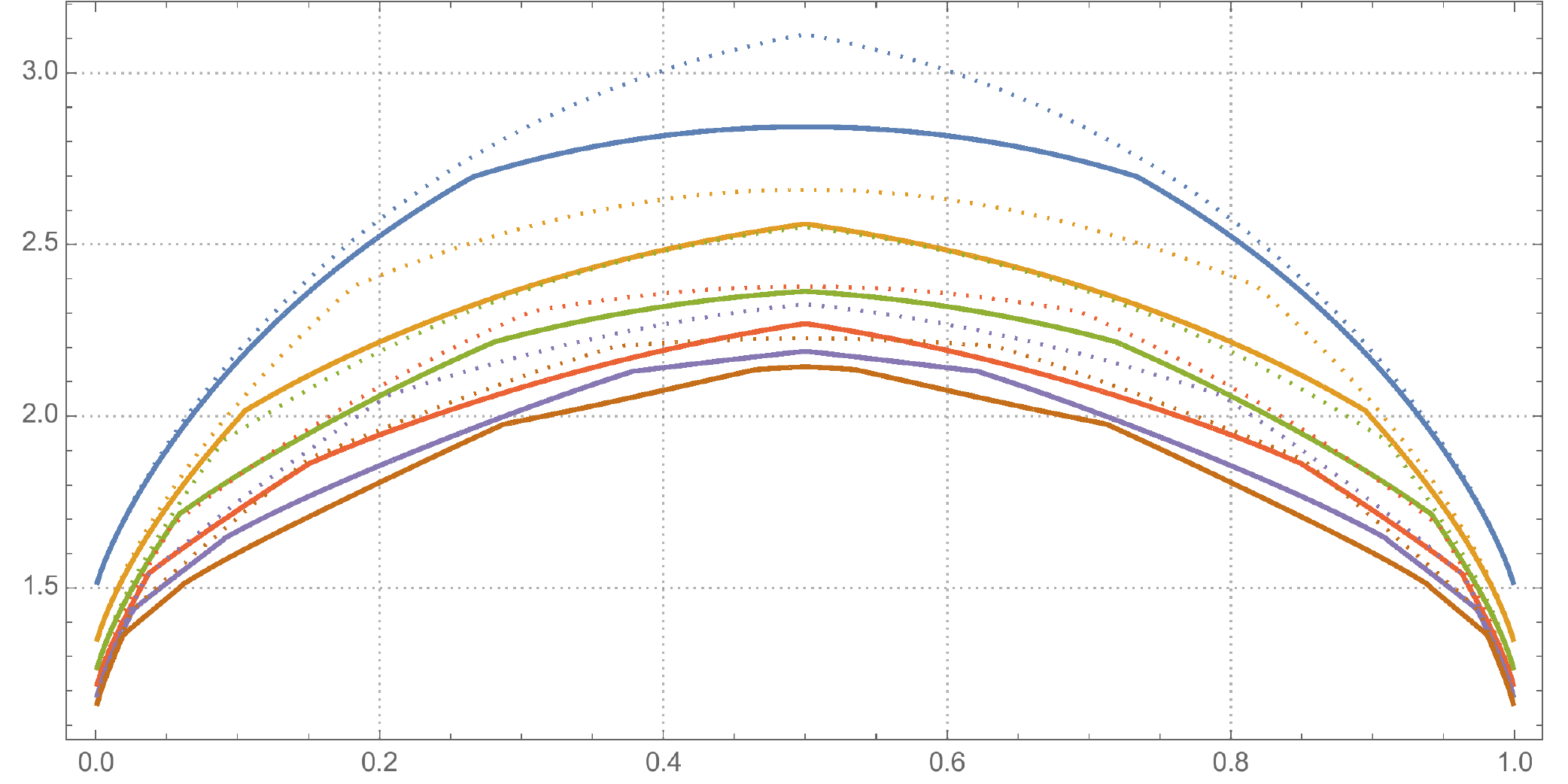}
	}%
	\caption{%
		$\mathit{se}(\alpha)$ for 
		SQS$k$ (thick) and (optimally biased) PROP$k$ (dotted) for 
		$k=2$ (blue), 
		$k=3$ (yellow), 
		$k=4$ (green), 
		$k=5$ (red),
		$k=6$ (purple), and
		$k=7$ (brown).%
	}
	\label{fig:sesquickselects-se-alpha}
\end{figure}

\section{Conclusion}
\label{sec:conclusion}

Despite the asymptotic optimality of the Floyd-Rivest algorithm,
practical implementations use Quickselect variants with a fixed-size sample.
Since they hence look very similar to sorting methods based on Quicksort,
it is tempting to copy optimizations that fair well in sorting blindly to the selection routines.
However, our results show that the similarities are misleading.
While multiway partitioning is vital in Quicksort for saving memory transfers~-- 
a cost measure of increasing relevance~--
more than two pivots are \emph{not} helpful in Quickselect.

Moreover, Quickselect offers a large potential for optimization
that has no counterpart in sorting whatsoever:
adapting the strategy to the (relative) sought rank $\alpha=\frac mn$.
The biased proportion-from-$k$ variants of single-pivot Quickselect
proposed in~\cite{MartinezPanarioViola2010}
minimize the number of \emph{comparisons}; 
in terms of scanned elements, however, Sesquickselect~-- 
a novel combination of single-pivot and dual-pivot Quickselect~-- 
outperforms proportion-from-$k$ significantly.

In the limit for large sample sizes $k\to\infty$,
Sesquickselect converges to the Floyd-Rivest algorithm.
This limit is ``degenerate'' in that we always choose \emph{two} pivots
(Sesquickselect only uses a single pivot for extreme $\alpha$),
and that the middle segment has size $o(n)$ (in expectation).
In that case, also the savings of dual-pivot partitioning over 
its simulation by two binary partitioning rounds are negligible.

For practical sample sizes, one cannot rely on this connection
to design a good selection method, though~-- 
unlike for single-pivot variants,
mimicking Floyd-Rivest too closely can result in performance much worse 
than non-adaptive Quickselect.
We need analyses that explicitly take the effect of fixed-size samples into account; 
such are initiated in this article.

\subsection{Future Work}

We had to leave many interesting questions about Sesquickselect open;
some are not even known for single-pivot Quickselect.
\begin{itemize}[noitemsep]
\item 
How fast do the costs converge to the optimum as $k$ grows?
Only an upper bound for median-of-$k$ seems known~\cite[Thm.\,4]{Gruebel1999}.
\item 
In \wref{fig:sesquickselects}, 
the number of intervals seems to grow linearly with~$k$;
can we avoid the use of many different versions in adaptive methods
while still achieving (close to) optimal costs?
\item
Do the theoretical improvements translate to faster running time?
Preliminary explorations were promising although the relative 
improvements are small.
\item What is the order of the second term / the speed of convergence in the asymptotic expansion
of the costs for fixed quantiles?
\item How does adaptive sampling affect the variance, higher moments or full distribution of costs?
	Some results for PROP$k$ are shown in~\cite{KnofRoesler2012}.
\item Does adaptive sampling also improve the number of symbol comparisons?
\end{itemize}

\ifsiam{}{

\clearpage
\appendix

\onecolumn
\normalsize


\numberwithin{algorithm}{section}
\numberwithin{table}{section}
\numberwithin{figure}{section}

\part*{Appendix}
\pdfbookmark[0]{Appendix}{}
\ifsiam{}{%
	\manualmark%
	\markleft{\mytitle}%
	\markright{Appendix}
}
\section{Index of Notation}
\label{app:notations}

In this appendix, we collect the notations used in this work.

\newlength\notationwidth
\setlength\notationwidth{9em}

\subsection{Generic Mathematical Notation}
\begin{notations}[\notationwidth]
\notation{$\N$, $\N_0$, $\Z$, $\Q$, $\R$, $\C$}
	natural numbers $\N = \{1,2,3,\ldots\}$, 
	$\N_0 = \N \cup \{0\}$,
	integers $\Z = \{\ldots,-2,-1,0,1,2,\ldots\}$,
	rational numbers $\Q = \{p/q : p\in\Z \wedge q\in\N  \}$,
	real numbers $\R$, and complex numbers $\C$.
\notation{$\R_{>1}$, $\N_{\ge3}$ etc.}
	restricted sets $X_\mathrm{pred} = \{x\in X : x \text{ fulfills } \mathrm{pred} \}$.
\notation{$0.\overline 3$}
	repeating decimal; $0.\overline3 = 0.333\ldots = \frac13$; \\
	numerals under the line form the repeated part of 
	the decimal number.
\notation{$\ln(n)$, $\ld(n)$}
	natural and binary logarithm; $\ln(n) = \log_e(n)$, $\ld(n) = \log_2(n)$.
\notation{$\vect x$}
	to emphasize that $\vect x$ is a vector, it is written in \textbf{bold;}\\
	components of the vector are not written in bold: $\vect x = (x_1,\ldots,x_d)$;\\
	unless stated otherwise, all vectors are column vectors.
\notation{$\lim_{x\to a^+}$, $\lim_{x\to a^-}$}
	directed limits; 
	$\lim_{x\to a^+ f(x)}$ is $\lim_{n\to\infty} f(a + \epsilon_n)$ and 
	$\lim_{x\to a^- f(x)}$ is $\lim_{n\to\infty} f(a - \epsilon_n)$
	for a strictly positive sequence $(\epsilon_n)_{n\in\N}$ 
	with $\lim_{n\to\infty} \epsilon_n = 0$.
\notation{$X$}
	to emphasize that $X$ is a random variable it is Capitalized.
\notation{$[a,b)$}
	real intervals, the end points with round parentheses are excluded, 
	those with square brackets are included.
\notation{$[m..n]$, $[n]$}
	integer intervals, $[m..n] = \{m,m+1,\ldots,n\}$;
	$[n] = [1..n]$.
\notation{$[\text{stmt}]$, $[x=y]$}
	Iverson bracket, $[\text{stmt}] = 1$ if stmt is true, $[\text{stmt}] = 0$ otherwise.
\notation{$\|\vect x\|_p$}
	$p$-norm;
	for $\vect x\in\R^d$ and $p\in\R_{\ge1}$ we have 
	$\|\vect x\|_p = \bigl(\sum_{r=1}^{d} |x_r|\bigr)^{1/p}$.
\notation{$\|\vect x\|_\infty$}
	$\infty$-norm or maximum-norm; 
	for $\vect x\in\R^d$ we have $\|\vect x\|_\infty = \max_{r=1,\ldots,d} |x_r|$.
\notation{$\vect x+1$, $2^{\vect x}$, $f(\vect x)$}
	element-wise application on vectors; $(x_1,\ldots,x_d) + 1  = (x_1+1,\ldots,x_d+1)$ and 
	$2^{\vect x} = (2^{x_1},\ldots,2^{x_d})$; 
	for any function $f:\C\to\C$ write $f(\vect x) = (f(x_1),\ldots,f(x_d))$ etc.
\notation{$\total{\vect x}$}
	``total'' of a vector; for $\vect x = (x_1,\ldots,x_d)$, 
	we have $\total{\vect x} = \sum_{i=1}^d x_i$.
\notation{$\harm n$}
	$n$th harmonic number; $\harm n = \sum_{i=1}^n 1/i$.
\notation{$\Oh(f(n))$, $\pm\Oh(f(n))$, $\Omega$, $\Theta$, $\sim$}
	asymptotic notation as defined, \eg, by \cite[Section A.2]{Flajolet2009};
	$f=g\pm\Oh(h)$ is equivalent to $|f-g| \in \Oh(|h|)$.
\notation{$\Gamma(z)$}
	the gamma function, $\Gamma(z) = \int_0^\infty t^{z-1}e^{-t} \, dt$.
\notation{$\BetaFun(\alpha_1,\ldots,\alpha_d)$}
	$d$-dimensional beta function
	$\BetaFun(\alpha_1,\ldots,\alpha_d) = \prod_{i=1}^d \Gamma(\alpha_i)/ \Gamma(\alpha+\beta)$.
\notation{$a^{\underline b}$, $a^{\overline b}$}
	factorial powers notation of Graham et al.~\cite{ConcreteMathematics}; \\
	``$a$ to the $b$ falling resp.\ rising.''
\notation{$h(x)$}
	the binary base-$e$ entropy function
	$h(x) = -x\ln(x) - (1-x) \ln(1-x)$.
\notation{$\poly{a_n,\ldots,a_0}$}
	abbreviation for a polynomial with given coefficients;
	$\poly{a_n,\ldots,a_0} = a_n \nu^n+a_{n-1}\nu^{n-1}+\cdots+ a_1 \nu + a_0$.
	(used in \wref{app:constants-sqs}).
\end{notations}

\subsection{Stochastics-related Notation}
\begin{notations}[\notationwidth]
\notation{$\Prob{E}$, $\Prob{X=x}$}
	probability of an event $E$ resp.\ probability for random variable $X$ to
	attain value $x$.
\notation{$\E{X}$}
	expected value of $X$; I write $\E{X\given Y}$ for the conditional expectation
	of $X$ given $Y$, and $\Eover X{f(X)}$ to emphasize that expectation is taken 
	\wrt random variable~$X$.
\notation{$X\eqdist Y$}
	equality in distribution; $X$ and $Y$ have the same distribution.
\notation{$\indicatornobraces{E}$, $\indicator{X\le 5}$}
	indicator variable for event $E$, \ie, $\indicatornobraces{E}$ is $1$ if $E$
	occurs and $0$ otherwise;
	$\{X\le 5\}$ denotes the event induced by the expression $X \le 5$.
\notation{$\uniform(a,b)$}
	uniformly in $(a,b)\subset\R$ distributed random variable. 
\notation{$\uniform[a..b]$}
	discrete uniformly in $[a..b]\subset\Z$ distributed random variable. 
\notation{$\betadist(\alpha,\beta)$}
	beta distributed random variable with shape parameters $\alpha\in\R_{>0}$ and $\beta\in\R_{>0}$.
	$X\eqdist\betadist(\alpha,\beta)$ is equivalent to $(X,1-X)\eqdist\dirichlet(\alpha,\beta)$.
\notation{$\dirichlet(\vect \alpha)$}
	Dirichlet distributed random variable;
	$\vect \alpha \in \R_{>0}^d$.
\notation{$\binomial(n,p)$}
	binomial distributed random variable with $n\in\N_0$ trials and success probability $p\in[0,1]$;
	$X\eqdist \binomial(n,p)$ is equivalent to $(X,n-X)\eqdist\multinomial(n;p,1-p)$.
\notation{$\multinomial(n,\vect p)$}
	multinomially distributed random variable; 
	$n\in\N_0$ and $\vect p \in [0,1]^d$ with $\total{\vect p} = 1$.
\notation{$\betaBinomial(n,\alpha,\beta)$}
	beta-binomial distributed random variable;
	$n\in\N_0$, $\alpha,\beta\in\R_{>0}$;
	$X\eqdist \betaBinomial(n,\alpha,\beta)$ is equivalent to
	$(X,n-X) \eqdist \dirichletMultinomial(n;\alpha,\beta)$.
\notation{$\dirichletMultinomial(n,\vect \sigma)$}
	Dirichlet-multinomial distributed random variable;
	$n\in\N_0$, $\vect\sigma\in\R_{>0}^s$.
\end{notations}

\subsection{Notation for our Analysis}

Some notations for the analysis are exemplified in \wpref{fig:notations}.

\begin{notations}[\notationwidth]
\notation{$n$}
	length of the input array, \ie, the input size.
\notation{$m$}
	the rank to select, $m\in[1..n]$.
\notation{$\alpha$}
	relative sought rank, $\alpha = m/n \in [\frac1n,1] \subset [0,1]$.
\notation{$s$, $s(\alpha)$}
	number of segments, $s\ge2$;
	determines the number of pivots to be $s-1$.
	For adaptive methods, $s$ depends on $\alpha$.
\notation{$\vect t$, $\vect t(\alpha)$}
	pivot-sampling parameter;
	$\vect t \in \N^{s}$;
	see \wref{fig:notations}.
	For adaptive methods, $\vect t$ depends on $\alpha$.
\notation{$k$, $k(\alpha)$}
	sample size $k\in \N_{\ge s-1}$; 
	$k+1 = \total{(\vect t + 1)}$.
	For adaptive methods, $k$ depends on $\alpha$ (via $\vect t$).
\notation{$\vect{\tau}$}
	quantiles vector for sampling, $\vect \tau = \frac{\vect t+1}{k+1}$;
	we have $\E{D_\ell} = \tau_\ell$.

\notation{$\mathcal F$, $\mathcal C$, $\mathcal{SE}$, $\mathcal{WA}$}
	cost measures; $\mathcal F$ is the generic placeholder for an arbitrary cost measure,
	$\mathcal C$ means number of \underline comparisons,
	$\mathcal{SE}$ means number of \underline scanned \underline elements,
	$\mathcal{WA}$ means number of \underline write \underline accesses to the array.
\notation{$F_{n,m}$, $C_{n,m}$, $\mathit{SE}_{n,m}$, $\mathit{WA}_{n,m}$}
	(random) costs in the respective cost measure to select the $m$th smallest out of $n$
	elements.
\notation{$M_n$, $F_{n,M_n}$, $F_n$}
	(random) costs (in the respective cost measure) to select a uniform random rank 
	$M_n \eqdist \uniform[1..m]$ out of $n$ elements.
\notation{$f(\alpha)$, $c(\alpha)$, $\mathit{se}(\alpha)$, $\mathit{wa}(\alpha)$}
	leading-term coefficient of $\E{F_{n,m}}$ for $n\to\infty$ and $m/n \to \alpha$:
	\(f(\alpha) \wwrel= \lim_{{n\to\infty, \frac mn \to\alpha}} \frac{\E{F_{n,m}}}{n}\);\\
	similarly for specific cost measures.
\notation{$\overline f$, $\overline c$, $\overline{\mathit{se}}$, $\overline{\mathit{wa}}$}
	leading-term coefficient of grand average
	\(
			\bar f 
		\wwrel= 
			\lim_{n\to\infty} \frac{\E{F_{n,M_n}}}{n} 
		\wwrel=
			\int_0^1 f(\alpha) \,d\alpha
	\);
	similar for specific cost measures.
\notation{$A_{\mathcal F}(n,m)$, $A_{\mathcal C}(n,m)$, $A_{\mathcal{SE}}(n,m)$, $A_{\mathcal{WA}}(n,m)$}
	(random) costs of the first partitioning step (in the resp.\ cost measure)
	when selecting the $m$th smallest out of $n$ elements.\\
	(Not directly a function of $m$, but indirectly for adaptive methods).
\notation{$a_{\mathcal F}$, $a_{\mathcal F}(\alpha)$}
	leading-term coefficient of $\E{A_{\mathcal F}(n,m)}$ for $n\to\infty$ with $m/n\to\alpha$.
	(Not directly a function of $m$, but indirectly for adaptive methods).
\notation{$P_1,\ldots,P_{s-1}$; $P_0$, $P_{s}$}
	(random) values of chosen pivots in the first partitioning step, \\
	ordered by value
	$0\le P_1\le P_2 \le \cdots \le P_{s-1}\le 1$; \\
	$P_0 \ce 0$ and $P_{s} \ce 1$ are used for notational convenience.
\notation{$R_1,\ldots,R_{s-1}$; $R_0$, $R_{s}$}
	(random) ranks of the pivots in the first partitioning step, \\
	$1\le R_1\le R_2 \le \cdots \le R_{s-1}\le n$; \\
	$R_0 \ce 0$ and $R_{s} \ce n+11$ are used for notational convenience.
\notation{$\vect J\in\N^{s}$}
	(random) vector of subproblem sizes for recursive calls;\\
	for initial size $n$, we have $\vect J \in \{0,\ldots,n-(s-1)\}^{s}$ with
	$\total{\vect J1} = n-(s-1)$.
\notation{$\vect I\in\N^{s}$}
	(random) vector of segment sizes;
	for initial size $n$, we have $\vect I \in \{0,\ldots,n-k\}^{s}$ with
	$\total{\vect I} = n-k$;\\
	$\vect J = \vect I + \vect t$ and $\vect I \eqdist \dirichletMultinomial(n-k,\vect t + 1)$;
	conditional on $\vect D$ we
	have $\vect I \eqdist \multinomial(n-k,\vect D)$.
\notation{{$\vect D\in[0,1]^{s}$}}
	(random) spacings of the unit interval $(0,1)$ induced by the pivots $P_1\,\ldots,P_{s-1}$, 
	\ie, $D_i = P_i - P_{i-1}$ for $1\le i \le s$;
	$\vect D \eqdist \dirichlet(\vect \sigma) \eqdist \dirichlet(\vect t + 1)$.
\notation{$H$, $H(\vect t)$}
	``const-entropy''
	$H(\vect t) = 1-\sum_{r=1}^s{\frac{\rfact{(t_r+1)}2}{\rfact{(k+1)}2}}$.
\end{notations}

\section{Proof of Hölder integration lemma}
\label{app:hölder}

In this appendix, we prove the error bound on the integral of
Hölder-continuous functions.

\begin{proof}{\wref{lem:hölder-intergral-bound}}
Let $C$ be the Hölder-constant of $f$.
We split the integral into small integrals over hypercubes of side length $\frac1n$
and use Hölder-continuity to bound the difference to the corresponding summand:
\begin{align*}
	&
	\mkern-50mu
		\Biggl|
		\int_{\vect x\in [0,1]^d} f(\vect x) \, d\vect x
		\wbin-
		\frac1{n^d} \!\! \sum_{\vect i \in [0..n-1]^d} \!\!\! f(\vect i / n)
		\Biggr|
\\	&\wwrel=
		\sum_{\vect i \in [0..n-1]^d} 
			\Biggl|
				\int_{\vect x \in [0,\frac1n]^d + \frac{\vect i}n} f(\vect x) \, d\vect x
				\bin- \frac{f(\vect i /n)}{n^d}
			\Biggr|
\\	&\wwrel=
		\sum_{\vect i \in [0..n-1]^d} 
				\int_{\vect x \in [0,\frac1n]^d + \frac{\vect i}n} 
				\bigl| f(\vect x) - f(\vect i /n)\bigr|
				\, d\vect x
\\	&\wwrel\le
		\sum_{\vect i \in [0..n-1]^d} 
				\int_{\vect x \in [0,\frac1n]^d + \frac{\vect i}n} 
				\sqrt dC \bigl\| \vect x - \tfrac{\vect i}n\bigr\|^h
				\, d\vect x
\\	&\wwrel\le
		\sqrt dC\sum_{\vect i \in [0..n-1]^d} 
				\int_{\vect x \in [0,\frac1n]^d + \frac{\vect i}n} 
				\bigl(\tfrac1n\bigr)^h
				\, d\vect x
\\	&\wwrel=
		\sqrt dCn^{-h} \int_{\vect x \in [0,1]^d} 1 \, d\vect x
\\	&\wwrel=
		\Oh(n^{-h}).
\end{align*}
\end{proof}

\section{Proof of Local Limit Law}
\label{app:local-limit-beta}

In this appendix, we give the proof for the local 
limit law of the Dirichlet-multinomial distribution.
It is a straight-forward generalization of the computation
given in \cite[Lemma~2.38]{Wild2016}, but
we include it for a self-contained presentation.

\begin{proof}[\wref{lem:local-limit-law-dirichlet-multinomial}]
	Let $\vect z\in(0,1)^s$ with $\total{\vect z} = 1$ be arbitrary and write 
	$\vect i=\vect i(\vect z)=\lfloor \vect z(n+1)\rfloor \in [0..n]^s$ with $\total{\vect i} = n$.
	We note for reference that for $\ell\in[s]$ and any constant $c\ge 0$ 
	\begin{align*}
			\bigl(\tfrac{i_\ell}n\bigr)^c
		&\wwrel=
			\biggl(\frac{\lfloor z_\ell(n+1) \rfloor}{n}\biggr)^{\!c}
		\wwrel=
			\bigl(z_\ell \pm \Oh(\tfrac1n)\bigr)^c
		\wwrel=
			z_\ell^c \bin\pm \Oh(\tfrac1n)
			,\qquad(n\to\infty).
	\numberthis\label{eq:floor-i-close-zn}
	\end{align*}
	Moreover, we use the following property of the gamma function
	\begin{align}
	\label{eq:gamma-function-quotient-rule}
			\frac{\Gamma(z+a)}{\Gamma(z+b)} 
		&\wwrel=
			z^{a-b} \pm \Oh(z^{a-b-1}),
			\qquad(z\to\infty), \quad(a,b \text{ constant}).
	\end{align}
	This follows from \href{http://dlmf.nist.gov/5.11#E13}{Equation~(5.11.13)} of the DLMF~\cite{DLMF}.
	With these preparations, we compute
	\begin{align*}
		&\mkern-18mu
			n^{s-1}\,\Prob{\ui{\vect I}{n}\! = \vect i}
	\\	&\wwrel=
			n^{s-1}\binom n{\vect{i}} \frac{\BetaFun(\vect i + \vect \alpha)}{\BetaFun(\vect\alpha)}
	\\	&\wwrel=
			\frac{n^{s-1}}{\BetaFun(\vect\alpha)} \cdot 
			\frac{\Gamma(n+1)}{\prod_{\ell=1}^s \Gamma(i_\ell+1)}\cdot
			\frac{\prod_{\ell=1}^s \Gamma(i_\ell+\alpha_\ell)}{\Gamma(n+\total{\vect\alpha})}
	\\	&\wwrel=
			\frac{n^{s-1}}{\BetaFun(\vect\alpha)} \cdot 
			\frac{\Gamma(n+1)}{\Gamma(n+\total{\vect\alpha})} \cdot
			\prod_{\ell=1}^s\frac{\Gamma(i_\ell+\alpha_\ell)}{\Gamma(i_\ell+1)}
	\\	&\wwrel{\eqwithref{eq:gamma-function-quotient-rule}}
			\frac{n^{s-1}}{\BetaFun(\vect\alpha)} \cdot 
			n^{1-\total{\vect\alpha}}\bigl( 1 \pm \Oh(n^{-1})\bigr) \cdot
			\prod_{\ell=1}^s \begin{cases*}
				1 & if $\alpha_\ell = 1$ \\
				i_\ell^{\alpha_\ell-1} \pm \Oh(i_\ell^{\alpha_\ell-2})
					& if $\alpha_\ell \ge 2$ 
			\end{cases*}
	\\[1ex]	&\wwrel=
			\frac{1}{\BetaFun(\vect\alpha)} \cdot 
			\bigl( 1 \pm \Oh(n^{-1})\bigr)\cdot 
			\prod_{\ell=1}^s \begin{dcases*}
				1 & if $\alpha_\ell = 1$ \\
				\bigl(\tfrac{i_\ell}n\bigr)^{\alpha_\ell-1} 
					\pm \Oh\Bigl(\smash{\underbrace{\bigl(\tfrac{i_\ell}n\bigr)^{\alpha_\ell-2}}_{\Oh(1)} n^{-1}}\Bigr)
				& if $\alpha_\ell \ge 2$
			\end{dcases*}
	\\[1ex]	&\wwrel{\eqwithref{eq:floor-i-close-zn}}
			\frac{1}{\BetaFun(\vect\alpha)} \cdot 
			\bigl( 1 \pm \Oh(n^{-1})\bigr) \cdot
			\prod_{\ell=1}^s \bigl( z_\ell^{\alpha_\ell-1} \pm \Oh(n^{-1})\bigr)
	\\	&\wwrel=
			\frac{\prod_{\ell=1}^s z_\ell^{\alpha_\ell - 1}}{\BetaFun(\vect\alpha)} 
			\wbin\pm \Oh(n^{-1}).
	\end{align*}
	This is exactly the density $f_D(\vect z)$ of the $\dirichlet(\vect\alpha)$ distribution.
\end{proof}

\section{Constants for Sesquickselect}
\label{app:constants-sqs}

In this appendix, we give the constants for Sesquickselect as functions in $\nu$.
With one exception, the expressions are rather unwieldy and we will have to use
some shorthand notation to state concisely.
Constant $C_2(\nu)$ is very simple;
for both comparisons and scanned elements we have $C_2(\nu)=2$.

In order to express the remaining constants, it will be convenient to introduce the following two 
auxiliary functions, namely,
\begin{align*}
		\Delta(\nu)
	&\wwrel\ce
		2\Biggl(60\ln(1-\nu)\nu^6-360\ln(1-\nu)\nu^5-140\nu^6
		+780\ln(1-\nu)\nu^4
\\*	&\wwrel\ppe\qquad{}
		+480\nu^5-840\ln(1-\nu)\nu^3-635\nu^4+504\ln(1-\nu)\nu^2
\\*	&\wwrel\ppe\qquad{}
		+428\nu^3-168\ln(1-\nu)\nu-156\nu^2+24\ln(1-\nu)
		+24\nu\Biggr),
\shortintertext{and}
		Q_a(\nu)
	&\wwrel\ce
		(\nu^4-4\nu^3+4\nu^2-2\nu+a)\cdot(1-\nu)^2\ln(1-\nu), \qquad (a\in\R).
\end{align*}
We will also use the shorthand $Q(\nu):= Q_{1/2}(\nu)$.

\begin{table}[b]
	\smaller[2]
\plaincenter{
	\begin{tabular}{l p{0.43\linewidth} p{0.43\linewidth}}
	\toprule
	  & \textbf{Comparisons} & \textbf{Scanned Elements} \\
	\midrule
	 $C'_1$   
	 	& $\poly{20,-120,260,-276,162,-52,7}$ 
	 	& $\poly{20,-120,260,-264,144,-40,4}$ 
	 	\\
	\midrule
	 $C'_3$ 
	 & 
		 $\begin{multlined}
		 12Q(\nu)(\ln\nu-\ln(1-\nu))\\
		 +\poly{6,16,-69,70,-24,-2,2}\ln(1-\nu) \\
		 +\poly{-26,92,-125,86,-33,6}\nu\ln\nu \\
		 +\poly{45,-124,121,-52,5,2}\nu
		   \end{multlined}$
	 &
		 $\begin{multlined}
		 48Q(\nu)(\ln\nu-\ln(1-\nu))\\
		 +\poly{90,-308,510,-560,408,-176,32} \ln(1-\nu) \\
		 +\poly{-110,380,-506,344,-132,24}\nu\ln\nu \\
		 +\poly{45,-40,-125,212,-136,32}\nu
		   \end{multlined}$
	  \\
	  \midrule
	 $C'_4$ 
	 &
		 $\begin{multlined}
		 -12Q(\nu)\ln(1-\nu)\\
		 +\poly{168,-956,2031,-2180,1317,-446,65}\cdot\\
		 \ln(1-\nu)\\
		 -\frac{20}{3}\nu^9+30\nu^8-\frac{170}{3}\nu^7\\
		 +\frac{1}{6}\poly{-1644,6792,-9409,6514,-2445,390}\nu
		     \end{multlined}$
	 &
		 $\begin{multlined}
		 -48Q(\nu)\ln(1-\nu)\\
		 +\poly{198,-956,1890,-2000,1236,-440,68}\cdot\\
		 \ln(1-\nu)\\
		 -\frac{20}{3}\nu^9+30\nu^8-\frac{170}{3}\nu^7\\
		 +\frac{1}{3}\poly{-456,2352,-3641,2756,-1146,204}\nu
		 \end{multlined}$
	\\ \midrule
	$C'_5$ 
	 &
	    $\begin{multlined}
	      228Q_{15/38}(\nu)\\
	      +\poly{-534,1828,-2415,1626,-591,90}\nu
	    \end{multlined}$
	 & 
	    $\begin{multlined}
	        192Q_{3/8}(\nu)\\
	        +\poly{-450,1540,-2034,1368,-492,72}\nu
	      \end{multlined}$
	 \\
	 \bottomrule
	 \end{tabular}
	}
 \caption{Constants for comparisons and scanned elements in Sesquickselect.}
 \label{tab:sesquickselect-constants}
 \end{table}

Moreover, we will use the notation $\poly{a_n,\ldots,a_0}$ to denote the polynomial of degree $n$ 
\[
a_n\nu^n +a_{n-1}\nu^{n-1}+\cdots+a_1\nu +a_0.
\]

With all these definitions at hand, the expressions for the constants $C'_i(\nu)=C_i(\nu)\cdot \Delta(\nu)$ 
are collected in \wref{tab:sesquickselect-constants}.

\needspace{5\baselineskip}
\section{Optimal Threshold for Sesquickselect}
\label{app:opt-sqs-nu}

In this appendix, we prove the existance of a unique optimal cutoff $\nu^*$.

\begin{proof}[\wref{thm:sqs-optimal-nu}]
Given a function $g$ in $[0,1]$ and a value $\nu$, $0\le \nu\le 1/2$, consider the following operator

\begin{align*}
		\bigl(T(g)\bigr)(\alpha)
	&\wwrel= 
		2\Biggl(
			-\alpha^3\int_\alpha^1\frac{g(x)}{x^4}\,dx+
			\alpha^2\int_\alpha^1\frac{g(x)}{x^3}\,dx
\\*	&\wwrel\ppe\quad{}
			+(1-\alpha)^3\int_0^\alpha\frac{g(x)}{(1-x)^4}\,dx
		\Biggr)
		,\qquad(\alpha < \nu),
\shortintertext{and}
		\bigl(T(g)\bigr)(\alpha)
	&\wwrel= 
		2\Biggl(
			\alpha^2\int_\alpha^1\frac{g(x)}{x^3}\,dx
			-\alpha^3\int_\alpha^1\frac{g(x)}{x^4}\,dx
\\*	&\wwrel\ppe\quad{}
			+(1-\alpha)^2\int_0^\alpha\frac{g(x)}{(1-x)^3}\,dx
			-(1-\alpha)^3\int_0^\alpha\frac{g(x)}{(1-x)^4}\,dx
\\*	&\wwrel\ppe\quad{}
			+\frac{(1-\alpha)^3}{3}\int_0^\alpha\frac{g(x)}{(1-x)^3}\,dx
			+\frac{\alpha^3}{3}\int_\alpha^1\frac{g(x)}{x^3}\,dx
		\Biggr)
		,\qquad(\nu\le\alpha\le1-\nu).
\end{align*}
Notice that the function $f_\nu$ for Sesquickselect satisfies $f_\nu=a_\mathcal{F}+T(f_\nu)$; it is important
to emphasize that the function $a_\mathcal{F}$ will depend on $\nu$ since we use single partitioning
if $\alpha<\nu$ or $\alpha>1-\nu$, whereas we use YBB partitioning if $\nu\le \alpha\le 1-\nu$.

To be more precise, the operator $T$ is defined piecewise: $T=[T_{1,\nu},T_{2,\nu}]$, and we can define it
so that for $h=T(g)$, we have $h_{1,\nu}=T_{1,\nu}(g_{1,\nu},g_{2,\nu})$ and
$h_{2,\nu}=T_{2,\nu}(g_{1,\nu},g_{2,\nu})$, with $h_{1,\nu}$ being the restriction of $h$
to $[0,\nu)\cup(1-\nu,1]$, $h_{2,\nu}$ the restriction of $h$ to $[\nu,1-\nu]$ and similarly
for $g_{1,\nu}$ and $g_{2,\nu}$.

Solving the fix-point equation $f = T(f)$ involves the same steps that we have followed to find the explicit
solution for Sesquickselect (the differential equations are identical). But we arrive here to the conclusion
that $f=0$ is the unique solution of $f = T(f)$.
(Note that the summand $a_{\mathcal F}$ is missing in the fix-point equation.)

Let $g_1(\nu)=\mathit{se}_{1,\nu}$ and $g_2(\nu)=\mathit{se}_{2,\nu}$. 
Both functions are strictly increasing in $(0,\frac12)$ 
since their derivatives with respect to $\nu$ are strictly positive. 
Moreover, we have
$g_1(0) = \frac53 < 2 = g_2(0)$ and $g_1(\frac12) \approx 3.112 > 2.910 \approx g_2(\frac12)$, 
so they cross at a single point $\nu=\nu^\ast$.
In other words, there exists a unique solution $\nu^\ast$ in $(0,\frac12)$
to the equation $g_1(\nu)=g_2(\nu)$.

Taking derivatives with respect to $\nu$ on both sides of 
$\mathit{se}_\nu=a_\mathcal{SE}+T(\mathit{se}_\nu)$ and
setting $\nu=\nu^\ast$, many terms cancel out because
$\mathit{se}_{1,\nu^\ast}(\nu^\ast) = \mathit{se}_{2,\nu^\ast}(\nu^\ast)$ 
and because of the symmetries
$\mathit{se}_{3,\nu}(\alpha) = \mathit{se}_{1,\nu}(1-\alpha)$ and 
$\mathit{se}_{2,\nu}(\alpha) = \mathit{se}_{2,\nu}(1-\alpha)$, 
so we conclude
\[
		\left.\frac{\partial se_\nu}{\partial\nu}\right|_{\nu=\nu^\ast}
	\wwrel=
			T\left(\left.\frac{\partial se_\nu}{\partial\nu}\right|_{\nu=\nu^\ast}\right),
\]
and hence, $\left.\partial se_\nu/\partial\nu\right|_{\nu=\nu^\ast}=0$ for any $\alpha\in[0,1]$.

Computing the second derivative of $\mathit{se}_\nu(\alpha)$ with respect to $\alpha$ and 
setting $\nu=\nu^\ast$
shows it is positive for all $\alpha$ in $(0,1)$, so we can conclude that $\nu^\ast$
is always a local minimum.
In fact, we can show that
\[
		\left.\frac{\partial^2 se_\nu}{\partial\nu^2}\right|_{\nu=\nu^\ast}
	\wwrel=
		\delta(\alpha)+T\left(\left.\frac{\partial^2 se_\nu}{\partial\nu^2}\right|_{\nu=\nu^\ast}\right),
\]
for some function $\delta(\alpha)$ which is strictly positive for all $\alpha\in[0,1]$. 
To be more precise,
$\delta(\alpha)=\delta_1(\alpha)$ if $\alpha\in[0,\nu^\ast)\cup(1-\nu^\ast,1]$ and
$\delta(\alpha)=\delta_2(\alpha)$
if $\nu^\ast\le \alpha\le 1-\nu^\ast$; both $\delta_1$ and $\delta_2$ are strictly positive in the
corresponding ranges of $\alpha$. 
Finally, $\delta>0$ for all $\alpha$ 
entails that $\left.\partial^2 \mathit{se}_\nu/\partial\nu^2\right|_{\nu=\nu^\ast}>0$ for all $\alpha$.

The limit values $\nu\to 0$ (YQS) and $\nu\to1/2$ (PROP2) are not minimum, hence to complete the proof we only
need to show that for any fixed value of $\alpha$, $\mathit{se}_\nu(\alpha)$ has no other local extrema in $(0,\frac12)$.
So let $\hat{\nu}$ be such that
$\left.\partial se_\nu/\partial\nu\right|_{\nu=\hat{\nu}}=0$ for some $\hat\alpha$.
Assume \withoutlossofgenerality that $\alpha\le \frac12$; then
if
$\hat\alpha<\hat{\nu}$ we assume that 
$\left.\partial \mathit{se}_{1,\nu}/\partial\nu\right|_{\nu=\hat{\nu}}=0$,
and if $\alpha > \nu$ then 
$\left.\partial \mathit{se}_{2,\nu}/\partial\nu\right|_{\nu=\hat{\nu}}=0$.
Since $\mathit{se}=a_\mathcal{SE}+T(\mathit{se})$ and
$\partial a_\mathcal{SE}/\partial\nu = 0$, we have
$\partial \mathit{se}_\nu/\partial\nu = \partial T(\mathit{se}_\nu)/\partial\nu$. 
Setting $\nu=\hat{\nu}$ on both sides,
we arrive at the conclusion that if 
$\left.\partial se_\nu/\partial\nu\right|_{\nu=\hat{\nu}}=0$ for some
$\hat\alpha$ then we must have
$\mathit{se}_{1,\hat{\nu}}(\hat{\nu}) = \mathit{se}_{2,\hat{\nu}}(\hat{\nu})$. 
This implies $\hat\nu = \nu^\ast$ since~-- as argued above~-- the latter is 
the unique value in $(0,\frac12)$ with 
$\mathit{se}_{1,\hat{\nu}}(\hat{\nu}) = \mathit{se}_{2,\hat{\nu}}(\hat{\nu})$.~
\end{proof}

\section{Proof of Convergence for Linear Ranks}
\label{app:convergence}

This appendix gives the details for the proof of \wref{thm:convergence}.

\subsection{Derivation of the integral equation}

We start with the distributional equation \wref{eq:dist-rec-general}
and take expectations on both sides.
\begin{align*}
		\E{F_{n,m}}
	&\wwrel=
		\E{A_{\mathcal F}(n,m)}
		\bin+ \sum_{\ell=1}^s \Eover[\Big]{\vect J}{ 
			\indicator{R_{\ell-1} < m < R_\ell} \cdot 
				\E{F_{J_\ell, m-R_{\ell-1}} \given \vect J}
		}
		.
\numberthis\label{eq:recurrence-Fnm-with-E}
\end{align*}
Unfolding the expectation and simplifying yields
\begin{align*}
		\sum_{\ell=1}^s \Eover[\Big]{\vect J}{ 
					\indicator{R_{\ell-1} < m < R_\ell} \cdot 
						\E{F_{J_\ell, m-R_{\ell-1}} \given \vect J}
				} \mkern-300mu
\\[-1ex]	&\wwrel=
		\sum_{j=1}^{n-1}
			\sum_{a=1}^{m}
			\Biggl( \sum_{\ell=1}^s \Prob{J_\ell = j \wedge m - R_{\ell-1} = a} \Biggr)
				\cdot \E{F_{j,a}}
\\	&\wwrel=
		\sum_{\underline r=0}^{m-1} \sum_{\overline r=\underline m+1}^{n+1}
			\Biggl( \sum_{\ell=1}^s \Prob{R_{\ell-1} = \underline r \wedge R_{\ell} = \overline r} \Biggr)
				\cdot \E{F_{\overline r - \underline r - 1,m - \underline r}}
	.
\end{align*}
(Note that the extreme values $0$ and $n+1$ for $\underline r$ resp.\ $\overline r$
are included to handle the outermost segments $\ell=1$ resp.\ $\ell=s$ and those with $1<\ell<s$
in uniform way. Eventually, these two cases will have to be distinguished, though,
since $R_1,\ldots,R_{s-1}$ are actual random variables, 
whereas $R_0 = 0$ and $R_{s+1} = n+1$ are deterministic values.)
We can express the joint probabilities
$\Prob{ (R_{\ell-1},R_{\ell}) = (\underline r , \overline r )}$
using the relations between $\vect R$ and $\vect J$ and using the aggregation property
of the Dirichlet-multinomial distribution
(see, \eg, \cite[\href{https://www.wild-inter.net/publications/html/wild-2016.pdf.html\#pf66}{Lemma 2.37}]{Wild2016}),
but we do not need an explicit expression for now.
We are interested in the limit $n,m\to\infty$ with $\frac mn \to \alpha \in (0,1)$,
\ie, we are selecting the $\alpha$-quantile of a large list.
Since we expect the overall costs to be asymptotically linear,
we divide the recurrence by $n$:
\begin{align*}
		\frac{\E{F_{n,m}}}n
	&\wwrel=
		\frac{\E{A_{\mathcal F}(n,m)}}n
		\bin+ \sum_{1\le \underline r < \overline r \le n}
			\frac{\overline r - \underline r - 1}{n} \times{}
\\*	&\wwrel\ppe{}\quad
			\Biggl( \sum_{\ell=1}^s \Prob[\big]{(R_{\ell-1},R_{\ell}) = (\underline r, \overline r) } \Biggr)
				\cdot \frac{\E{F_{\overline r - \underline r - 1,m - \underline r}}}{\overline r - \underline r - 1}
		.
\numberthis\label{eq:scaled-recurrence}
\end{align*}

\subsection{Proof of convergence}

\wref{thm:convergence} essentially claims that we can take 
the limit in \wref{eq:scaled-recurrence} to obtain a characterizing equation
for $f(\alpha)$.

\begin{proof}[\wref{thm:convergence}]
It suffices to prove the asymptotic approximation since it implies the existence of the limit.
We first connect \weqref{eq:general-integral-equation} (disregarding convergence issues)
to the normalized expected costs $\E{F_{n,m}} / n$
and then prove the claimed asymptotic approximation.

We start with the \textsl{ansatz} $\E{F_{n,m}} \sim f(\frac mn) n$ for a function $f:[0,1] \to \R$.
Inserting into \weqref{eq:scaled-recurrence} yields for $\frac mn \to \alpha$
\begin{align*}
		f(\alpha)
	&\wwrel=
		a_\mathcal F(\alpha) 
		\bin+ \sum_{\ell=1}^s \int_{\underline z=0}^\alpha \int_{\overline z = \alpha}^1 
			(\overline z- \underline z) 
			f_{(P_{\ell-1},P_\ell)}(\underline z, \overline z)\cdot
			f\Bigl(\frac{\alpha- \underline z}{\overline z- \underline z}\Bigr)
			\, d \overline z \, d \underline z
		,
\numberthis\label{eq:integral-equation-one-line}
\end{align*}
where
\begin{align*}\SwapAboveDisplaySkip
		f_{(P_{\ell-1},P_\ell)}(\underline z, \overline z)
	&\wwrel=
		\begin{dcases}
			\delta_0(\underline z) \cdot \frac{\overline z^{t_1} \, (1-\overline z)^{\underrightarrow{t_1}}}
				{\BetaFun(t_1+1,\underrightarrow{t_1}+1)},
				& \ell = 0; \\
			\frac{
				\underline z^{\underleftarrow{t_{\ell}}} \,
				(\overline z - \underline z)^{t_\ell} \, 
				(1-\overline z)^{\underrightarrow{t_{\ell}}}
			}
			{\BetaFun( \underleftarrow {t_\ell}+1, t_\ell+1, \underrightarrow{t_\ell}+1 )},
				&1 < \ell < s; \\
			\delta_1(\overline z) \cdot \frac{\underline z^{\underleftarrow{t_s}} \, (1-\overline z)^{t_s}}
				{\BetaFun(\underleftarrow{t_s}+1,t_s+1)} ,
				& \ell = s .
		\end{dcases}
\end{align*}
Here we write $\delta_c(z)$ for the \textsl{Dirac-delta function} at $c$,
the degenerate density of a random variable that deterministically has value $c\in\R$.
By separating the leftmost and rightmost cases and inserting $f_{(P_{\ell-1},P_\ell)}$,
we obtain the claimed integral equation for $f$.

Let now $\alpha\in(0,1)\setminus\mathcal A$ be fixed and set $m=\lceil\alpha n\rceil$.
We consider the quantity $r_{n,m} = \E{F_{n,m}} - f(\frac mn)n$.
Inserting the recurrence \wref{eq:recurrence-Fnm-with-E} for $\E{F_{n,m}}$ 
and the integral equation \wref{eq:integral-equation-one-line} for $f(\frac mn)$ yields
\begin{align*}
		r_{n,m}
	&\wwrel=
		b_r(n,m)\cdot n
		\bin+ 
		\sum_{\ell=1}^s \Eover*{\vect J}{
				\indicator{R_{\ell-1} < m < R_\ell} \cdot 
					r_{R_\ell-R_{\ell-1}-1, m-R_{\ell-1}}
		}
\numberthis\label{eq:recurrence-r-nm}
\shortintertext{with}
		b_r(n,m)
	&\wwrel=
		a_\mathcal F(\tfrac mn) \bin\pm\Oh(n^{-1}) - a_{\mathcal F}(\alpha)
\\*	&\wwrel\ppe\quad{}
		\bin+
		\sum_{\ell=1}^s \sum_{\underline r=0}^{m-1} \sum_{\overline r= m+1}^{n+1} \!\!
			\frac{\overline r - \underline r - 1}{n}\cdot
			\Prob{R_{\ell-1} = \underline r \wedge R_{\ell} = \overline r} 
				\cdot f\biggl(\frac{m - \underline r}{\overline r - \underline r - 1}\biggr)
\\*	&\wwrel\ppe\quad{}
		\bin-
		\sum_{\ell=1}^s \int_{\underline z=0}^\alpha \int_{\overline z = \alpha}^1 
			(\overline z- \underline z) 
			f_{(P_{\ell-1},P_\ell)}(\underline z, \overline z)\cdot
			f\Bigl(\frac{\alpha- \underline z}{\overline z- \underline z}\Bigr)
			\, d \overline z \, d \underline z.
\end{align*}
The outline of the remainder of the proof is to first show that $b_r(n,m)$ is small,
namely $b_r(n,m) = \Oh(n^{-2h/3})$, uniformly in $\alpha$.
For that we show that the difference above is essentially the difference of an integral 
and the corresponding Riemann sum. Since the integrand is sufficiently smooth,
we can bound the absolute value of this difference, but some case distinctions will be necessary for that.
Finally, we show that small $b_r(n,m)$ implies that also $r_{n,m}$ is small thus completing the proof.

First of all, not that 
for $\alpha\notin\mathcal A$, we have for large enough $n$ that $\frac mn \in I_v$ iff $\alpha \in I_v$.
Hence also 
$a_{\mathcal F}(\frac mn) = a_{\mathcal F}(\alpha)$ for large $n$ and the two terms cancel.
So remains to compare the sums and integrals.
Setting $\underline z = \underline r / n$ and $\overline z = \overline r / n$ 
we find by Hölder-continuity of $f$ inside the $I_v$
for $\frac{\alpha-\underline z}{\overline z - \underline z} \notin\mathcal A$ that
\(
	f(\frac{m - \underline r}{\overline r - \underline r - 1})
	=
	f(\frac{\lceil\alpha n\rceil - \underline z n}{\overline z n - \underline z n - 1})
	=
	f(\frac{\alpha-\underline z}{\overline z - \underline z}) \pm \Oh(n^{-h})
\)
for $h\in(0,1)$ the Hölder exponent of $f$.
Moreover, 
\(
	\Prob{R_{\ell-1} = \underline r \wedge R_{\ell} = \overline r}
	=
	\Prob{R_{\ell-1}/n = \underline z \wedge R_{\ell}/n = \overline z}
	=
	n^{-2} f_{P_{\ell-1},P_\ell}(\underline z, \overline z)
	\pm \Oh(n^{-3})
\)
by \wref{lem:local-limit-law-dirichlet-multinomial}.
This yields (separating the left- and rightmost call cases)
\begin{align*}
		b_r(n,m)
	&\wwrel=
		\frac1{n} \sum_{m < \overline r \le n}
			\overline z\cdot  f_{P_1}(\overline z) 
			\cdot f\biggl(\frac{\alpha}{\overline z}\biggr)
		\wwbin-
			\int_{\overline z = \alpha}^1 \overline z\cdot  f_{P_1}(\overline z) 
						\cdot f\biggl(\frac{\alpha}{\overline z}\biggr)\, d\overline z
\\*	&\wwrel\ppe{}
		\bin+\frac1{n} \sum_{1 \le \underline r < m}
			(1-\underline z) f_{P_s}(\underline z) 
			\cdot f\biggl(\frac{\alpha-\underline z}{1 - \underline z}\biggr)
		\wwbin-
			\int_{\underline z = 0}^\alpha (1-\underline z) f_{P_s}(\underline z) 
					\cdot f\biggl(\frac{\alpha-\underline z}{1 - \underline z}\biggr) \, d \underline z
\\*	&\wwrel\ppe{}
		\bin+
		\sum_{\ell=2}^{s-1}\; \frac1{n^2}\!\! \sum_{1\le \underline r < m < \overline r \le n} \!\!\!
			(\overline z-\underline z) f_{P_{\ell-1},P_\ell}(\underline z,\overline z) 
			\cdot f\biggl(\frac{\alpha-\underline z}{\overline z - \underline z}\biggr)
\\*	&\wwrel\ppe\qquad{}
		\bin-
		\sum_{\ell=2}^{s-1} \int_{\underline z=0}^\alpha \int_{\overline z = \alpha}^1 
			(\overline z- \underline z) 
			f_{(P_{\ell-1},P_\ell)}(\underline z, \overline z)\cdot
			f\biggl(\frac{\alpha- \underline z}{\overline z- \underline z}\biggr)
			\, d \overline z \, d \underline z
\\*	&\wwrel\ppe{}
		\wbin\pm\Oh(n^{-1} + n^{-h}).
\end{align*}
Unfortunately, the function 
\[
	g_\alpha:[0,\alpha]\times[\alpha,1]\to[0,1]\quad \text{with}\quad
	g_\alpha(\underline z, \overline z) \rel= \frac{\alpha-\underline z}{\overline z - \underline z}
	,
\]
which appears as argument of $f$ 
(with $\underline z = 0$ in the first line and $\overline z = 1$ in the second line)
varies rapidly if $\overline z - \underline z$ becomes small,
which causes the sum and integral to differ significantly in that region.
Formally, the partial derivatives of $g_\alpha$ approach $-\infty$ as 
$(\underline z,\overline z) \to (\alpha,\alpha)$.
If we exclude this point and its vicinity, however,
$g_\alpha$ is well-behaved,
in particular $g_\alpha$ restricted to $D_\alpha(n) = [0,\alpha-n^{-1/3}]\times[\alpha+n^{1/3}]$
has partial derivatives bounded by $\frac14 n^{1/3}$ in absolute value.
This implies the following Lipschitz-condition for $g_\alpha$
\begin{align*}
		\forall (u,v) , (u',v') \in D_\alpha(n)
		\wrel:
		\bigl| g_\alpha(u,v) - g_\alpha(u',v') \bigr|
		\wwrel\le n^{1/3}\cdot \bigl\| (u,v) - (u',v') \bigr\|_2.
\numberthis\label{eq:g-alpha-lipschitz}
\end{align*}
We will therefore consider $D_{\alpha}(n)$ and 
$\overline D_{\alpha}(n) = [0,\alpha]\times[\alpha,1] \setminus D_\alpha(n)$
separately.

Let us first consider $\overline D_{\alpha}(n)$. 
We will bound the sum and integral in isolation and show that their absolute contribution
is small, so even though we cannot bound their difference well, the overall contribution of
this regime is insignificant.
The integrals resp.\ sums restricted to that domain become 
(we show the middle case $\ell=2,\ldots,s-1$; other outermost ones are similar)
\begin{align*}
	&\wwrel\ppe{}
		\Biggl| \mkern45mu
		\tfrac 1{n^2} \cdot {}\mkern-75mu \sum_{m-n^{2/3}\le \,\underline r\, < m < \,\overline r\, \le m+n^{2/3}} 
			\mkern-50mu
			(\overline z-\underline z) f_{P_{\ell-1},P_\ell}(\underline z,\overline z) 
			\cdot f\biggl(\frac{\alpha-\underline z}{\overline z - \underline z}\biggr)
		\Biggr|
\\*	&\wwrel\ppe{}
		\bin+
		\Biggl|\int_{\underline z=\alpha-n^{-1/3}}^\alpha \int_{\overline z = \alpha}^{\alpha+n^{-1/3}}
			(\overline z- \underline z) 
			f_{(P_{\ell-1},P_\ell)}(\underline z, \overline z)\cdot
			f\biggl(\frac{\alpha- \underline z}{\overline z- \underline z}\biggr)
			\, d \overline z \, d \underline z\,
		\Biggr|\;;
\intertext{%
	$f$ is continuous on the bounded domains $I_v$ and hence absolutely bounded
	and $f_{P_{\ell-1},P_\ell}$ is a polynomial, hence continuous and likewise bounded, 
	so this is
}
	&\wwrel\le
		\Biggl| \mkern45mu
		\tfrac 1{n^2} \cdot {}\mkern-75mu \sum_{m-n^{2/3}\le \,\underline r\, < m < \,\overline r\, \le m+n^{2/3}} 
			\mkern-50mu
			(\overline z-\underline z) \cdot \Oh(1) 
		\Biggr|
		\bin+
		\Biggl|
		\int_{\underline z=\alpha-n^{-1/3}}^\alpha \int_{\overline z = \alpha}^{\alpha+n^{-1/3}}
			(\overline z- \underline z) \cdot \Oh(1)
			\, d \overline z \, d \underline z\,
		\Biggr|,
\intertext{and with $\overline z-\underline z = \Oh(n^{-1/3})$}
	&\wwrel\le
		\Oh\bigl(n^{-1/3}\bigr)\cdot 
		\frac{n^{2/3}\cdot n^{2/3}}{n^2}
		\wwbin+
		\Oh\bigl(n^{-1/3}\bigr)\bin\cdot n^{-1/3}\cdot n^{-1/3}
\\	&\wwrel=
		\Oh(n^{-1}).
\end{align*}
Note that this bound holds uniformly in $\alpha$.
For the outermost cases, $\ell=1$ and $\ell=s$,
a similar result holds, but the final error bound is only $\Oh(n^{-2/3})$ 
since the fraction of summands in $\overline D_\alpha(n)$ is only $n^{2/3} / n$.

For the remaining region of integration, we split the large integral into a sum of small rectangles.
Denoting the integrand by 
\[
		I(\underline z,\overline z) 
	\wwrel= 
		(\overline z-\underline z) f_{P_{\ell-1},P_\ell}(\underline z,\overline z) 
			\cdot f\biggl(\frac{\alpha-\underline z}{\overline z - \underline z}\biggr)
\]
and setting $\underline z = \underline r / n$ and $\overline z = \overline r / n$ as before, 
we obtain
\begin{align*}
	&\wwrel\ppe{}
		\Biggl|
		\frac1{n^2}\!\! \sum_{\substack{1\le \,\underline r\, \le m-n^{2/3}-1\\ 
							m+n^{2/3}+1 \le\, \overline r\, \le n}} \mkern-15mu
			(\overline z-\underline z) f_{P_{\ell-1},P_\ell}(\underline z,\overline z) 
			\cdot f\biggl(\frac{\alpha-\underline z}{\overline z - \underline z}\biggr)
\\*	&\wwrel\ppe\quad{}
		\bin-
		\int_{u=0}^{\alpha-n^{-1/3}} \mkern-8mu \int_{v = \alpha+n^{-1/3}}^1 
			(v- u) 
			f_{(P_{\ell-1},P_\ell)}(u,v)\cdot
			f\biggl(\frac{\alpha- u}{v-u}\biggr)
			\, d v \, d u
		\,\Biggr|
\\	&\wwrel\le
		\sum_{\substack{1\le \,\underline r\, \le m-n^{2/3}-1\\ m+n^{2/3}+1 \le\, \overline r\, \le n}} \Biggl|
		\frac1{n^2} \cdot
			I(\underline z, \overline z)
		\bin-
		\int_{u=\underline z - \frac1n}^{\underline z} \int_{v = \overline z-\frac1n}^{\overline z}
			I(u,v)
			\, d v \, d u
		\,\Biggr|
\\	&\wwrel=
		\sum_{\substack{1\le \,\underline r\, \le m-n^{2/3}-1\\ m+n^{2/3}+1 \le\, \overline r\, \le n}}
		\int_{u=\underline z - \frac1n}^{\underline z} \int_{v = \overline z-\frac1n}^{\overline z}
		\bigl|
			I(u,v) - 
			I(\underline z, \overline z)
		\bigr|
		\, d v \, d u
		.
\numberthis\label{eq:diff-integrals-tame-region}
\end{align*}
To bound the difference between the integrands, we use that $I$ is ``smooth'';
some care is needed to trace errors.
We first collect properties of the involved functions.
$f$~is Hölder-continuous inside the $I_v$'s, so there is a constant $C_f$ with
\begin{align*}
		\forall v\in[d] \;\forall x,y\in I_v\wbin:
		\bigl|f(x) - f(y)\bigr| \wrel\le C_f |x-y|^h.
\numberthis\label{eq:f-Hölder}
\end{align*}
It follows that $f$ is bounded in each $I_v$ and hence their finite union:
\begin{align*}
		\forall x\in[0,1]
	&\wbin:
		|f(x)| \wrel\le \hat f
		.
\numberthis\label{eq:f-bounded}
\end{align*}
Now denote by $h(u,v) = (v-u)f_{P_{\ell-1},P_\ell}(u,v)$ (so $I(u,v) = h(u,v) f(g_\alpha(u,v))$).
$h(u,v)$ is a polynomial in $u$ and $v$, hence Lipschitz-continuous and bounded on a compact domain.
\begin{align*}
		\forall (u,v),(u',v') \in \{ (x,y) \in [0,1]^2: x\le y \}
	&\wbin:
			\bigl| h(u,v) - h(u',v') \bigr| 
		\wrel\le 
			C_P \| (u,v) - (u',v') \|_2
			,
\numberthis\label{eq:h-lipschitz}
\\
		\forall (u,v)\in \{ (x,y) \in [0,1]^2: x\le y \}
	&\wbin:
			\bigl| h(u,v) \bigr| 
			\wrel\le \hat h
			.
\numberthis\label{eq:h-bounded}
\end{align*}
Now consider any $(\underline r,\overline r) \in [1..m-n^{2/3}-1] \times [m+n^{2/3}+1,n]$,
so that there is a single $\alpha$-interval $I_v$ with
$g_\alpha\bigl( [\underline z-\frac1n,\underline z]\times[\overline z-\frac1n, \overline z] \bigr) \subset I_v$.
Call such a pair $(\underline r,\overline r)$ \emph{pure}.
(Its associated integration region does not contain a boundary of adaptivity intervals.)
For any pure $(\underline r, \overline r)$ and 
$(u,v) \in [\underline z-\tfrac1n,\underline z]\times[\overline z-\tfrac1n, \overline z]$
now holds
\begin{align*}
		\bigl| I(u,v) - I(\underline z, \overline z) \bigr|
	&\wwrel\le
		\phantom{{}\bin+{}}
		\Bigl|
			  h(u,v) f\bigl(g_\alpha(u,v)\bigr) 
			- h(u,v) f\bigl(g_\alpha(\underline z,\overline z)\bigr)
		\Bigr| 
\\*	&\wwrel\ppe{}
		\bin+ \Bigl|
			  h(u,v) f\bigl(g_\alpha(\underline z,\overline z)\bigr)
			- h(\underline z,\overline z) f\bigl(g_\alpha(\underline z,\overline z)\bigr)
		\Bigr|
\\	&\wwrel{\relwithtworefs{eq:f-Hölder}{eq:h-lipschitz}\le}
		h(u,v)\cdot C_f \bigl| g_\alpha(u,v) - g_\alpha(\underline z,\overline z) \bigr|^h
		\bin+
		f\bigl(g_\alpha(\underline z,\overline z)\bigr)\cdot C_P\|(u,v) - (\underline z,\overline z)\|
\\	&\wwrel{\relwiththreerefs{eq:f-bounded}{eq:h-bounded}{eq:g-alpha-lipschitz}\le}
		\hat h\cdot C_f \cdot n^{h/3}\bigl\| (u,v) - (\underline z,\overline z) \bigr\|^h
		\bin+
		\hat f\cdot C_P\|(u,v) - (\underline z,\overline z)\|
\\	&\wwrel=
		\Oh(n^{h/3-h}).
\numberthis\label{eq:diff-integrands}
\end{align*}
Using this in \weqref{eq:diff-integrals-tame-region} yields
\begin{align*}
	&
	\Biggl|	
		\sum_{\substack{1\le \,\underline r\, \le m-n^{2/3}-1\\ m+n^{2/3}+1 \le\, \overline r\, \le n}} 
			\frac1{n^2} \cdot
				I(\underline z, \overline z)
			\bin-
			\int_{u=\underline z - \frac1n}^{\underline z} \int_{v = \overline z-\frac1n}^{\overline z}
				I(u,v)
				\, d v \, d u
			\,
	\Biggr|
\\	&\wwrel\le
		\sum_{\substack{1\le \,\underline r\, \le m-n^{2/3}-1\\ m+n^{2/3}+1 \le\, \overline r\, \le n}} 
		\int_{u=\underline z - \frac1n}^{\underline z} \int_{v = \overline z-\frac1n}^{\overline z}
		\begin{dcases*}
		\Oh(n^{-h \cdot2/3}) & if $(\underline r,\overline r)$ pure \\
		\hat h \cdot \hat f & otherwise
		\end{dcases*}
\\	&\wwrel\le
		\Oh(n^{-h \cdot2/3})\wbin+ \Oh\biggl( \frac{\#\text{non-pure } (\underline r, \overline r)}{n^2}\biggr)
		.
\numberthis\label{eq:diff-integrals-tame-region-part2}
\end{align*}
To bound the second term, we observe that by \wref{eq:g-alpha-lipschitz}, 
$g_\alpha\bigl( [\underline z-\frac1n,\underline z]\times[\overline z-\frac1n, \overline z] \bigr)\subset [x,x+\sqrt2n^{-2/3}]$
for some $x\in[0,1]$,
\ie, $g_\alpha$ can only span a very narrow range inside one patch of the integral.
Indeed, $g_\alpha(u,v)$ is (weakly) decreasing in both $u$ and $v$, so $x = g_\alpha(\underline z, \overline z)$.
Hence $(\underline r,\overline r)$ is pure if 
$[g_\alpha(\underline z, \overline z),g_\alpha(\underline z, \overline z) + \sqrt2n^{-2/3}] \cap \mathcal A = \emptyset$.

We also have to show that $g_\alpha$ is not ``too flat'', so that we cannot have many patches
containing the same interval boundary. More precisely, for any given $x\in[0,1]$,
we claim there are only $\Oh(n^{4/3})$ pairs $(\underline r, \overline r)$ with 
$x\in g_\alpha\bigl( [\underline z-\frac1n,\underline z]\times[\overline z-\frac1n, \overline z] \bigr)$.
We distinguish two cases.
\begin{enumerate}
\item $x \le \frac12$: Let $v$ be fixed. Then we have $g_\alpha(u,v) \in [x,x+\epsilon]$
	iff $u \in \bigl[\frac{\alpha-vx}{1-x},\frac{\alpha-v(x+\epsilon)}{1-(x+\epsilon)}\bigr]$.
	The size of this range for $u$ is
	\begin{align*}
			\Biggl|\frac{\alpha-vx}{1-x}\bin-\frac{\alpha-v(x+\epsilon)}{1-(x+\epsilon)}\Biggr|
		&\wwrel=
			\epsilon \cdot \biggl|\frac{v-\alpha}{(1-x)(1-x-\epsilon)}\biggr|
		\wwrel\le
			\epsilon \cdot \frac{1}{\frac12 (\frac12-\epsilon)}
		\wwrel\le
			5\epsilon
	\end{align*}
	for $\epsilon\le 0.1$.
	Setting $\epsilon=\sqrt2 n^{-2/3}$, this holds for $n\ge \sqrt 2 \cdot 10^{3/2}\approx 45$.
	This means, for every choice of $\overline r$, there are $\Oh(n^{1/3})$ choices for $\underline r$
	for which $(\underline r,\overline r)$ might be non-pure.
\item $x>\frac12$: As above, but we fix $u$ and find that the range for $v$ is
	\begin{align*}
			\Biggl|\frac{\alpha-u(1-x)}{x}\bin-\frac{\alpha-u(1-(x+\epsilon))}{x+\epsilon}\Biggr|
		&\wwrel=
			\epsilon\biggl|\frac{\alpha-u}{x(x+\epsilon)}\biggr|
		\wwrel\le
			4\epsilon.
	\end{align*}
\end{enumerate}
Summing over the constant number of interval boundaries in $\mathcal A$ yields
$\Oh\bigl( \frac{\#\text{non-pure } (\underline r, \overline r)}{n^2}\bigr) = \Oh(n^{-2/3})$.

The same bound as in \weqref{eq:diff-integrals-tame-region-part2} follows for the integrals
with $\ell=1$ and $\ell=s$: \weqref{eq:diff-integrands} holds also for that case
by setting $\underline z = 0$ resp.\ $\overline z = 1$, and the number of non-pure pairs
turns into the number of non-pure values of $\underline r$ resp.\ $\overline r$,
which can be bounded in the same way as above.

Finally combining the results for the regimes $D_\alpha(n)$ and $\overline D_\alpha(n)$,
we obtain the bound on $b_r(n,m) = \Oh(n^{-h\cdot 2/3})$

\medskip\noindent
Hence the error terms $r_{n,m}$ fulfill the same recurrence as $\E{F_{n,m}}$,
but with a smaller toll function of order $\Oh(n^{\epsilon})$ with $\epsilon=1-2h/3 \in (\frac13,1)$.
To finish the proof, we consider $\hat r_n = \max_{m} r_{n,m}$.
Since $r_{n,m} \le \hat r_n$, the claim follows if we can show that 
$\hat r_n = \Oh(n^{\epsilon})$.
From the recurrence~\wref{eq:recurrence-r-nm} for $r_{n,m}$, we obtain
\begin{align*}
		\hat r_n
	&\wwrel=
		\Oh(n^\epsilon) + 
		\max_{1\le m\le n} \sum_{\ell=1}^s \Eover[\Big]{\vect J}{
			\indicator{R_{\ell-1} < m < R_\ell} \cdot r_{R_\ell-R_{\ell-1}-1,m-R_\ell-1}
		}
\\	&\wwrel\le
		\Oh(n^\epsilon) + 
		\max_{1\le m\le n} \sum_{\ell=1}^s \Eover[\Big]{\vect J}{
			\indicator{R_{\ell-1} < m < R_\ell} \cdot \hat r_{J_\ell}
		}
\\	&\wwrel\le
		\Oh(n^\epsilon) + 
		\Eover[\Big]{\vect J}{\max_{1\le \ell\le s}  \hat r_{J_\ell}}
\\	&\wwrel=
		\Oh(n^\epsilon) + 
		\hat r_{\E{\max J_\ell}}
		.
\end{align*}
Clearly, this means we lose a constant fraction of $n$ in each step and
the first toll function will dominate.
More formally, we have
\begin{align*}
		\E[\Big]{\max_{1\le \ell\le s}J_\ell}
	&\wwrel=
		\sum_{j=\lceil n/s\rceil}^n
		j\cdot \Prob{\max J_\ell = j}
\\	&\wwrel\le
		\sum_{j=\lceil n/s\rceil}^n \,j\cdot \sum_{\ell=1}^s\Prob{J_\ell = j}
\\	&\wwrel{\eqwithref[r]{lem:local-limit-law-dirichlet-multinomial}}
		n\cdot \frac1n\sum_{j=\lceil n/s\rceil}^n
		\frac jn\sum_{\ell=1}^s \bigl(f_{P_\ell}(j/n) \pm \Oh(n^{-1})\bigr)
\\	&\wwrel{\eqwithref[r]{lem:hölder-intergral-bound}}
		n\cdot \underbrace{\sum_{\ell=1}^s \int_{1/s}^1 z f_{P_\ell}(z) \, dz}_{\equalscolon \rho \;<\; 1} 
		\wwbin\pm \Oh(1)
		.
\end{align*}
The constant is $\rho < 1$ since integrating from $0$ to $1$ would yield exactly $1$.
So we have $\hat r_n \le \hat r_{\rho n} + \Oh(n^\epsilon)$, and it is easy to prove by induction
that $\hat r_n = \Oh(n^\epsilon)$.
\end{proof}

}

\bibliography{quicksort-refs.bib}

\end{document}